# Identification of the Chromophores in Prussian blue

*Li Musen,[1,2][†] Robin Purchase,[3][†] Parvin Safari,[4] Martyna Judd,[3] Emily C. Barker, [4,5] Jeffrey R. Reimers,[1,2] Nicholas Cox,[3] Paul J. Low,[4] and Elmars Krausz[3]*

1 International Centre for Quantum and Molecular Structures and Department of Physics, Shanghai University, Shanghai 200444, China.

2 University of Technology Sydney, School of Mathematical and Physical Sciences, Ultimo, NSW 2007, Australia.

3 Research School of Chemistry, Australian National University, Canberra, ACT 2601 Australia.

4 School of Molecular Sciences, University of Western Australia, Perth W6009, Australia

5 Centre for Microscopy, Characterisation and Analysis, University of Western Australia, Perth, WA 6009, Australia

[†]: equal first authors

Email: jeffrey.reimers@uts.edu.au, robin.purchase@anu.edu.au, paul.low@uwa.edu.au, nick.cox@anu.edu.au, elmars.krausz@anu.edu.au

**ABSTRACT:** Prussian blue was the world's first synthetic dye. Its structural, optical and magnetic properties have led to many applications in technology and medicine, and provide paradigms for understanding coordination polymers, framework materials and mixed-valence compounds. The intense red absorption of Prussian blue that characterises chemical and physical properties critical to many of these applications is now shown to arise from localised intervalence charge transfer transitions within two chromophoric variants (ligand isomers) of an idealised "dimer" fragment $\{(NC)_5Fe^{II}\}(\mu-CN)\{Fe^{III}(NC)_3(H_2O)_2\}$. This fragment is only available in modern interpretations of the material's crystal structure, with the traditional motif $\{(NC)_5Fe^{II}\}(\mu-CN)\{Fe^{III}(NC)_5\}$ shown not to facilitate visible absorption. Essential to the analysis is the demonstration, obtained independently using absorption and magnetic circular dichroism spectroscopies, that spectra of Prussian blues are strongly influenced by particle size and (subsequent) light scattering. These interpretations are guided and supported by density functional theory calculations (CAM-B3LYP), supplemented by coupled cluster and Bethe-Salpeter spectral simulations, as well as electron paramagnetic resonance spectroscopy of Prussian blue and a model molecular dimeric ion $[Fe_2(CN)_{11}]^{6-}$.

Prussian blue (Fig. 1) is best described as a mixed-valence, three-dimensional coordination polymer (CP) containing Fe(II) and Fe(III) ions linked by bridging cyanide ions. Although semantic, the CP terminology is preferred over designation as a metal-organic framework (MOF) as cyanide is not usually classified as an organic ligand. The first synthesis of Prussian blue is attributed to Diesback in 1706 for use as a commercial dye (1). It now appears on the World Health Organization Essential Medicines list, and is widely used in cosmetics as a photoabsorbent, in photothermal cancer treatments (2-7) and other applications in medical imaging and drug delivery (7). Beyond biomedical applications, the structural, electronic, optical and magnetic properties of Prussian blue make for an outstandingly versatile platform for materials and molecular science research (8). For example, Prussian blue is ferromagnetic below a Curie temperature of 5.5 K (9, 10) and related materials are ferrimagnetic at higher temperatures (11). It has inspired designs of molecular magnets (12-14), including photo magnets (15-17) undergoing fs-timescale dynamics post absorption (18). The coordination framework of Prussian blue also provides a basis for designs for high-temperature superconductors (19, 20), supercapacitors (21), electrochromic devices (22-24), spin devices (11, 25, 26), and sensor actuators (27), as well as materials that change phase when they transport charge (28). Traditional uses of Prussian blue have resulted in the order of 10,000 tonnes being synthesised annually. More recently, Prussian blue is being used as an electrode in sodium-ion batteries for large-scale energy storage applications (29-32), with 10-fold production increases underway (33).

The intense absorption of red light by Prussian blue that results in the characteristic ink-blue colour of the solid is critical to many of its uses and is the subject of modification in many of its analogues. The absorption peak near 1.7 eV (730 nm) was proposed by Robin to arise from localised intervalence charge transfer (IVCT) between Fe(II) and high-spin (34-37) Fe(III) ions present in the crystal (38, 39), and has been analysed using the adiabatic electron-transfer theory of Hush (40, 41). However, as a framework material, Prussian blue samples with a wide range of chemical compositions are known, and although many have similar spectra, the primary IVCT absorption band properties can vary somewhat depending on the method of synthesis and environment (2, 4, 41-44). This enables 'Prussian blues' to be prepared in which this characteristic 1.7 eV absorption peak is relatively narrow (38), or alternatively, present with a long tail to low energy that extends through the bio-transparent window. Any absorption in this low energy region, as distinct from wavelength dependent light scattering, is critical to the use of Prussian blue in photothermal cancer treatments (2-7).

Modern high-resolution crystal structure work demonstrates that Prussian blue is a CP, but with poor long-range order. Its structure is based on a short-range motif represented as $\{Fe^{II}\}_3\{Fe^{III}\}_4(CN)_{18}(H_2O)_6 \cdot m\text{KOH} \cdot n\text{H}_2\text{O}$ (Fig. 1a) in which $K^+$ ions, $OH^-$ ions, and water molecules occupy pores within an uncharged $\{Fe^{II}\}_3\{Fe^{III}\}_4(CN)_{18}(H_2O)_6$ framework (44-48). It is possible to replace the potassium ion by other cations, a feature that is employed in the battery applications indicated above, and to substitute the iron ions, leading to an extensive set of Prussian blue analogues, many of which may be established to show the same basic structural motif (47). Prussian blues are typically characterised into two groups: "soluble" materials in which Prussian-blue particles stay in suspension for long periods, and "insoluble" materials in which particles settle out on the laboratory timescale. Nevertheless, these labels are misleading as for all samples the particles are suspended, not dissolved, in aqueous solution. It is common to correlate soluble Prussian blue with samples that have relatively narrow optical absorption and relatively high $K^+$ content, and conversely insoluble samples with broader absorption bands and low $K^+$ content. Minimising $K^+$ content by synthetic control has been a research focus, but traces $K^+$ may be found even in carefully synthesised samples of 'insoluble' Prussian blue (44, 49).

The general framework structure of Prussian blues, $\{Fe^{II}\}_3\{Fe^{III}\}_4(CN)_{18}(H_2O)_6 \cdot m\text{KOH} \cdot n\text{H}_2\text{O}$, can be considered to be a modification of a simpler (but charged) cubic framework, $\{Fe^{II}Fe^{III}(CN)_6\}$ (Fig. 1b), from which one of every four $\{Fe^{II}(CN)_6\}$ units is absent, with the resulting vacant coordination site in the surrounding six $Fe^{3+}$ ions being occupied by water. Long-range ordering of the missing $Fe^{2+}$ units within the framework is poor, and as a result (charged) short-range framework motifs such as $[\{Fe^{II}\}_4\{Fe^{III}\}_4(CN)_{24}]^{4-}$ and $[\{Fe^{II}\}_2\{Fe^{III}\}_4(CN)_{12}(H_2O)_{12}]^{4+}$ need to be included in accurate crystallographic modelling (44). Preparation conditions affect the long-range ordering of the framework and the ionic composition and occupancy of the pores (*e.g.* $0 \leqslant m < 2$), and hence also affect the $Fe^{III}:Fe^{II}$ ratio, which has been shown to vary from 1.09 to 1.33 (44). Hence the materials referred to as "Prussian blue" show considerable variation in chemical composition and internal structural arrangement, morphology and particle size (44). Of particular concern to the design of materials based on the Prussian blue framework is whether these chemical and physical variations lead to fundamental changes to the chromophore responsible for red-light absorption, or else simply act as external perturbations to some more intrinsic functional unit.

The above generalised, but more nuanced, description of a defect-riddled structure of Prussian blue has been developed only in the last 15 years (44, 46-48), and as a consequence, many of these key features are not reflected in descriptions of this material made over the course of its 300-year history. Indeed, the first crystal structure of Prussian blue, from 1936, reported its structure as hydrated $K[Fe^{II}Fe^{III}(CN)_6]$ (Fig. 1b) (39). However, it was somewhat later realised that this model incorrectly predicted the density of Prussian blue and its analogues and was thus untenable (45). Accordingly, an 'insoluble' counter ion-free form of hydrated [{Fe$^{II}$}$_3${Fe$^{III}$}$_4$(CN)$_{18}$] was described (49, 50) in 1977 (Fig. 1a), and forms the basis of the modern understanding that all Prussian blues have missing {Fe$^{II}$(CN)$_6$} structural units.

From the first interpretations of the low energy electronic absorption of Prussian blue in terms of an IVCT transition, it has been assumed that the primary chromophore is essentially an isolated mixed-valence {Fe$^{II}$}(μ-CN){Fe$^{III}$} dimeric unit (38, 40). Implicit in this interpretation is that exciton couplings between neighbouring units are small. This hypothesis is supported by the observation (51, 52) of analogous IVCT transitions in mixed-valence *bis*-metal species in solution such as salts of [Fe$_2$(CN)$_{11}$]$^{6-}$. As only one type of dimeric motif, {(NC)$_5$Fe$^{II}$}(μ-CN){Fe$^{III}$(NC)$_5$} appeared within both the originally perceived {Fe$^{II}$Fe$^{III}$(CN)$_6$} framework, and as the more recent structural descriptions based on {Fe$^{II}$}$_3${Fe$^{III}$}$_4$(CN)$_{18}$(H$_2$O)$_6$ also embody this structure, it has remained assumed that this chromophore is the origin of the IVCT band in Prussian blue. Nevertheless, given that it is now known that all Prussian blues are derived from the {Fe$^{II}$}$_3${Fe$^{III}$}$_4$(CN)$_{18}$(H$_2$O)$_6$, framework, the mixed-valence dimeric chromophore {(NC)$_5$Fe$^{II}$}(μ-CN){Fe$^{III}$(NC)$_3$(H$_2$O)$_2$} must also be present in all Prussian blues. Consequently, it is unclear which species is responsible for the observed absorption and its variations, or indeed if the observed spectra exhibit contributions from *both* chromophores.

Since the optical properties of Prussian blues lie at the heart of the myriad of applications of these structures, any attempt at structure-property relationship-based design of new materials demands a more accurate understanding of the nature of the key chromophore(s). Indeed, the characteristic optical absorption properties of Prussian blue could be generated by: a single chromophore; multiple similar chromophores; multiple disparate chromophores; or from excitonic interactions between chromophores. As Prussian blue particle sizes and morphologies are variable, it is also important to discriminate between intrinsic light absorption by the chromophores, and light scattering arising from its macroscopic particulate properties.

To address these issues, four Prussian blues have been synthesised and characterised (Table 1). These materials were obtained using either 'direct' (D) or 'indirect' (I) methods (vide infra). For each type, the synthetic conditions were varied to prepare samples identifiable as 'soluble' (s) or 'insoluble' (i) Prussian blues; the four materials are hence named **D$_s$**, **D$_i$**, **I$_s$**, and **I$_i$**. Direct syntheses involve mixing aqueous solutions of ferric and ferrocyanide ions leading to rapid precipitation of the mixed-valence framework material, whilst indirect syntheses combine ferrous and ferrocyanide ions to make homovalent Berlin white, of general formula [{Fe$^{II}$}$_3$(CN)$_6$], followed by slow aerial oxidation to the mixed-valence Prussian blue. In addition, two previously synthesised materials (**P**) are considered: that made by Robin (38), who synthesised a soluble Prussian blue with extremely small particle size, **P$_s$**, and that made by Buser et al. (49), who synthesised an insoluble material $P_i$ with large average particle size. The six materials considered display different compositions, particle sizes, apparent solubility, and spectra.

In this study, particle-sizing experiments on the synthesised materials were performed using laser diffraction (PSLD) and scanning electron microscopy (SEM). Absorption and magnetic-circular dichroism (MCD) spectra, the latter being relatively insensitive to light scattering of the sample but responsive to intrinsic absorption, have been recorded for these samples. To aid spectral interpretation, absorption and MCD spectra were recorded for the model mixed-valence dimer anion $[Fe_2(CN)_{11}]^{6-}$ (as a mixed $Na^+$ / $K^+$ salt) in solution, along with the application of X-band electron-paramagnetic resonance (EPR) spectroscopy and hyperfine sublevel correlation (HYSCORE) spectroscopy, to determine the valance, and associated ligand coordination (κ-CN vs κ-NC) and spin state, of the metal ions. These experiments resolve issues concerning the relative importance of light scattering and light absorption to the spectra of Prussian blues, and indicate a minimum number of identifiable internal chromophoric components within the framework.

A variety of computation methods have then been applied to interpret the optical spectra of Prussian blue and its model compound and identify the chemical nature of the chromophores. Density-functional theory (DFT) is widely used to interpret the physical and electronic structure, and electronic absorption spectra of ions and materials, including mixed-valence structures such as those found in Prussian blue (53-59). Notably, a band gap for the fragment $K[Fe^{II}Fe^{III}(CN)_6]$ (Fig. 1b) of 2.31 eV has been predicted from calculations with the HSE06 functional (58, 60). HSE06 has been the functional of choice for spectral calculations of materials (61), and the observation of an IVCT transition at 1.7 eV, assuming that the lowest-energy transition is allowed, supports the notion that $\{(NC)_5Fe\}(\mu-CN)\{Fe^{III}(NC)_5\}$ is the dominant chromophore. In contrast, multiconfigurational DFT methods (56) have predicted a much higher value of 3.8 eV for the $K[Fe^{II}Fe^{III}(CN)_6]$ band-gap. Also, empirical "DFT+U" methods, adjusted to equate the calculated bandgap for $\{(NC)_5Fe\}(\mu-CN)\{Fe^{III}(NC)_5\}$ to the observed 1.7 eV absorption maximum, have predicted (54) that $\{Fe^{II}\}_3\{Fe^{III}\}_4(CN)_{18}(H_2O)_6$ should have a bandgap of only some 0.6 eV, some 1.1 eV *lower* in energy than the parent structure obtained using the same level of theory.

These previous calculations have therefore variously attributed the observed spectra of Prussian blue samples to a range of chromophores. Unfortunately, none of these calculations enable a realistic treatment of the asymptotic potential of materials (62-74), and are thus all unlikely to make realistic *a priori* estimates of IVCT band gaps, exciton binding energies, and optical transition energies (53, 54, 58). Recently, the CAM-B3LYP functional (68, 69), that was designed explicitly to treat charge transfer transitions (64, 66, 67, 70, 72, 75), was made available in the VASP computational package (73, 76, 77). This Coulomb-attenuated functional has been demonstrated to reduce errors in calculated exciton binding energies fivefold compared to HSE06 (73, 74), with average and maximum errors in calculated bandgaps reduced by a factor of 2.5. This functional also avoids catastrophic failures associated with incorrect ground-state or excited-state identification (74). It therefore provides a significantly improved method towards the investigation of Prussian blue. We also apply the *ab initio* Bethe-Salpeter (78) (BSE) method to calculate spectra, including the properties of many excitonic states into the calculations. These results, combined with HSE06 calculations, as well as CAM-B3LYP, HSE06, and *ab initio* coupled-cluster based EOM-CCSD (79) and SAC-CI (80) calculations for the model compound and model material components help assign the observed optical transitions to chromophores, as identified in the most recent crystal structure(s).

The results are summarised as 11 deductions depicting key features of the spectral interpretation and modelling of Prussian blue.

**Results and Discussion**

**Synthesis.** Details of the syntheses (44, 49) of $D_s$, $D_i$, $I_s$, and $I_i$ are given in Supporting Information Section 1 and summarised in Table 1. A small quantity of the historical sample $P_i$ was provided by Andreas Ludi from the 1976 preparation (49).

**Particle sizing.** PSLD sample-size distributions for $D_s$, $D_i$, $I_s$, $I_i$ and $P_i$ are shown in Supporting Information Figure S1. Particle sizes of aggregates of $D_s$ and $D_i$ produced by the rapid precipitation characteristic of the 'direct' approach were centred around 60 – 80 μm, whereas those for $I_s$ and $I_i$ crystallised over many weeks during the 'indirect' synthesis had smaller sizes centred around 30 μm. The original 1976 sample ($P_i$) featured particle sizes centred at 10 μm. In each case, particles with sizes down to 1 μm were observed, but no smaller particles were detected down to the detection limit of 0.01 μm. The diversity of the shapes and sizes of the particles is highlighted in the SEM results shown in Supporting Information Figure S2 and Table S1. The indirect synthesis led to regular cubic-shaped crystalline particles, with relatively narrow size distributions (Table S1), whilst those from the direct methods gave irregular-shaped aggregates with a greater size distribution.

**Correlation between synthesis strategy, composition, particle size and shape, solubility, and absorption energy and bandwidth.** For the two historical samples ($P_s$ and $P_i$) and the four freshly synthesised ones ($D_s$, $D_i$, $I_s$, and $I_i$), Table 1 compares the synthetic strategy used

with potassium composition, particle sizing, apparent "solubility", and generic absorption properties.

The expectation that small changes in synthetic conditions result in samples characterisable as "soluble" and "insoluble" based on their rate of settling was met, as can be visualised in the Supporting Information Video. However, there was poor correlation between solubility and $K^+$ composition, with low composition materials displaying both soluble and insoluble forms, as indeed is also the case for high-$K^+$ materials. Also, no correlation between average particle size (SI) and stability of the suspension was found. Rather the aqueous suspensions of the regular cubic samples prepared by the indirect method, regardless of the designation as 'soluble' ($I_s$) or 'insoluble' ($I_i$), settled more quickly from solution, pointing to a lower zeta potential for these well-crystallised samples. The more irregular particles prepared by direct methods ($D_s$, $D_i$) settled over longer time courses, indicating a reduced tendency to aggregation in suspension and higher zeta potentials.

Concerning the width and energy of the IVCT band, on average the insoluble samples absorbed at lower energy with broader bands (Table 1), as per expectations. Nevertheless, the wide range of properties observed for each type indicates that solubility cannot be reliably predicted based upon absorption properties alone. These results lead to

- Deduction 1: Apparent "solubility" and "insolubility" is controlled more by the shapes and surface charge of the Prussian blue particles than by particle size or composition and is poorly anticipated based upon spectral properties.

**Details of IVCT absorption spectra of Prussian blue.** To illustrate the spectral variations that have been observed for Prussian blue samples, Fig. 2 compares the spectrum for soluble sample $P_s$ reported (38) by Robin in 1962, with spectra recorded here for suspensions of $D_s$, $D_i$, $I_s$, and $I_i$ in 1 cm pathlength cells. Robin's sample was prepared in such a way as to make extremely small particles in solution, facilitating accurate measurements of the intrinsic absorption. Note that Fig. 2 shows the bandshape function $A(\nu)/\nu$, which reflects the intrinsic absorption properties of the materials; as the IVCT band energy is low and the band is wide in comparison to it, consideration of raw absorption $A(\nu)$ presents a distorted view of the bandshape, its centre and width. Of particular note, in Fig. 2a the absorption of $I_i$ in the 0.5 – 1 eV region is much broader than that for $P_s$, suggestive of the presence of a second spectroscopic transition, as calculations have mooted (54).

Robin's work (38) is important as only his unique synthetic protocol yielded a material $P_s$ in which all particles were small enough to transmit light, hence providing the only true measure of the molar extinction of the material (9000 $M^{-1}$ $cm^{-1}$). This result tells that the *total* path length of light through *all* crystals in the light path through his 1 cm cell was just 180 nm. Hence only the smaller particles present in the $D_s$, $D_i$, $I_s$, and $I_i$ materials can contribute to absorption spectra. Larger particles may scatter incident light out of the solid angle seen by the detector in the spectrometer, or else be so strongly absorptive as to lead to no measurable transmission. Particles of 180 nm or less in size were too few to be detected by the PSLD (from suspensions stabilised by addition of 0.01% v/v sodium hexametaphosphate) or SEM measurements, and the *average* sizing reported in Table 1 therefore may not be a useful measure of the particles responsible for the observed spectroscopic properties.

The light scattering contribution to the absorption spectra of the set of Prussian blue samples was demonstrated by collecting sequential absorption spectra and attributing them to contributions from light scattering and intrinsic absorption (Supporting Information Section S3). The spectra observed for the direct-synthesis materials $D_s$ and $D_i$ showed little temporal variation, whereas for the samples produced by indirect synthesis, $I_s$ and $I_i$, the light-scattering component was found to decrease significantly with time, whilst the intrinsic absorption only narrowed a little (Figs. 2c and 2d, respectively).

The effect of light scattering on the spectral features of the Prussian blue samples is illustrated in Fig. 2e in which spectra after the longest standings for $D_s$, $D_i$, $I_s$, and $I_i$ are compared to each other, showing the raw values of $A/\nu$ measured and their fits to the Gaussian-absorption and light-scattering functions. The extracted absorption components of the spectra of $I_s$ and $I_i$ are also shown in the figure (dashed lines). These indicate variations in band centre $h\nu_v$ and full width at half maxim (FWHM) (Table 1) compared to the analogous materials made by direct syntheses $D_s$ and $D_i$, respectively. These results lead to the following deductions

- Deduction 2: a critical feature controlling the apparent energy and width of the IVCT band is light scattering arising from particles larger than the wavelength of light in suspension;
- Deduction 3: there is no evidence that an electronic absorption-band system exists at ca. 1 eV below the intense IVCT transition;
- Deduction 4: there is no evidence suggesting that significant chromophore variation occurs that can be correlated with solubility, composition, or synthetic approach.

The spectral fits in Fig. 2e show small deviations from Gaussian band shapes. These could arise from the presence of multiple, similar, chromophores within the crystal unit cell, exciton coupling, or excitation of crystal phonon modes accompanying the transition, with Gaussian bandshapes being favoured by crystal disorder. Excitation of phonons with energies exceeding the measurement temperature convert band profiles from Gaussian to Huang-Rhys form.

However, the observed deviation from a Gaussian form is the opposite to that expected, eliminating this as a possibility for explaining the observed bandshape.

**Prussian blue MCD spectra.** Absorption and MCD spectra measured at 1.8 K in 50:50 water/glycerol of the directly synthesised samples of soluble Prussian blue $D_s$ and insoluble Prussian blue $D_i$ are shown in Fig. 3, where they are compared to spectra of $P_i$ taken in polyvinyl alcohol (PVA) at 10 K for absorption in a bubble-free flow tube and at 1.8 K for MCD. The low temperature absorption spectra of $D_s$ and $D_i$ samples collected on the MCD spectrometer are similar to those recorded in water at room temperature (Fig. 2b), but the bands are 25% wider than those reported in Table 1 owing to the different solvation environment used in the MCD measurements.

The MCD intensity of a purely *linearly* polarised transition such as an idealised IVCT transition in Prussian blue is negligible. Nevertheless, MCD has been observed in molecular mixed valence compounds as a result of an interference mechanism between the linearly *z*-polarised intervalence process and a weaker single ion process that has *x* and/or *y* polarisation (81). The MCD seen in Prussian blue may arise from a related mechanism.

The MCD spectra for $D_i$, $D_s$, and $P_i$ shown in Fig. 3 are similar and do not reflect the variations in absorption that are also shown in the figure. These observations independently confirm Deductions 3 – 4 as there is no evidence suggesting a lower-energy transition, and no obvious qualitative dependence of the chromophore on the synthetic preparation. They also confirm Deduction 2 as MCD spectra are insensitive to light scattering, with the absorption spectral tails previously attributed to light scattering having no obvious counterpart in the MCD spectra.

A significant new feature revealed in the MCD spectra (Fig. 3) is the appearance of two apparent spectral components separated by ca. 0.6 eV. Saturation data for the MCD has been measured and analysed (Supporting Information Section S4) that shows uniform characteristics at all wavelengths, indicating that the two chromophores are similar. Also, the MCD analysis reflects well-known properties of the bulk magnetisation such as fast onset below the Curie temperature, and paramagnetic behaviour above the Curie temperature, again indicating chromophore similarity. This leads to

- <u>Deduction 5:</u> The IVCT band contains internal structure, pointing to at least two similar but distinguishable chromophores in all Prussian blues.

**Characterisation of the Prussian blue model mixed-valence molecular ion $[Fe_2(CN)_{11}]^{6-}$ in aqueous solution.** The $[Fe_2(CN)_{11}]^{6-}$ ion was synthesised as a model for the spectroscopic properties of Prussian blue (51), although the coordination environments at the Fe(III) sites of this model and in Prussian blue are not directly comparable (vide infra). The absorption spectrum of this bimetallic mixed-valence ion, measured as the $K_{4.7}Na_{1.3}[Fe_2(CN)_{11}]$ salt in aqueous solution, shows an intense localised IVCT transition (Fig. 4a) and higher-energy absorption features in good agreement with known spectra (51) (Fig. S5). Given the π-accepting nature of the C terminus of the cyanide ligand, the $[Fe_2(CN)_{11}]^{6-}$ ion is usually represented (51) by the localised valence structure $[\{(NC)_5Fe^{II}\}(\mu\text{-}CN)\{Fe^{III}(CN)_5\}]^{6-}$ (Fig. 1(c)), rather than the linkage isomer $[\{(NC)_5Fe^{III}\}(\mu\text{-}CN)\{Fe^{II}(CN)_5\}]^{6-}$. Given the strong-field nature of the cyanide ligand, the hexacyanoferrous ion can be safely assumed to be low-spin, but to the best of our knowledge, the spin state of the Fe(III) ion with the $(CN)\{Fe^{III}(CN)_5\}$ ligand sphere has not yet been determined. The intent here is to fully characterise the valence structure, magnetic properties and optical absorption spectra of this well-defined bimetallic mixed-valence model system to provide a control system for experimental data interpretation, as well as tests for computational models.

In Fig. 4a, the observed spectral bandshape $\epsilon/\nu$ of $[Fe_2(CN)_{11}]^{6-}$ is fitted using a single-mode Huang-Rhys model. The fit is of high quality and reveals an origin (zero-phonon line) energy of $h\nu_{00} = 0.33$ eV (3800 nm), a vertical excitation energy of $h\nu_v = 0.86$ eV (1450 nm), and a transition moment of $M = 0.88$ eÅ. The MCD spectrum of this mixed-valence model predominantly exhibits a single feature, suggesting that the IVCT transition embodies a single chromophore (Fig. 3), in contrast to Prussian blue.

The X-band EPR spectrum of $[Fe_2(CN)_{11}]^{6-}$ in frozen solution at 10 K is shown in Fig. 4b, where it is modelled using an EasySpin *pepper* simulation (82). This is interpreted as a spin $S = ½$ system dominated by $g$ anisotropy (see Supporting Information Section S6). The model compound therefore contains low-spin $Fe^{III}$, in contrast to the high-spin $Fe^{III}$ found in Prussian blue (9-11, 34-37). To identify the coordination of the $Fe^{III}$ ion on which the spin is located, X-band HYSCORE spectra were recorded (Supporting Information Section S6). The valence structure (Fig. 1c) is confirmed to be well represented as $[\{(NC)_5Fe^{II}\}(\mu\text{-}CN)\{Fe^{III}(CN)_5\}]^{6-}$ in line with expectations from ligand field theory, as well as analogies with other systems (83, 84). Further, the results show that the unpaired spin occupies the $(d_{xz}, d_{yz})$ orbital pair (defined with $z$ oriented along the $Fe^{II} - CN - Fe^{III}$ axis) and hence the ground state is $^2E$ in the $C_{4v}$ point group. A Jahn-Teller distortion leads to a breaking of the degeneracy of the $^2E$ state, such that the ground state becomes orbitally non-degenerate. Overall, the results indicate that

- <u>Deduction 6:</u> The valence structure of the $[Fe_2(CN)_{11}]^{6-}$ ion is $[\{(NC)_5Fe^{II}\}(\mu\text{-}CN)\{Fe^{III}(CN)_5\}]^{6-}$ with a low-spin, Jahn-Teller-distorted $^2E$ ground state that is characterised by an intense, low-energy, IVCT band.

**Modelling the structure and spectroscopic properties of the $[Fe_2(CN)_{11}]^{6-}$ model ion in solution.** Calculations were performed on the model ion using the CAM-B3LYP density functional and the SAC-CI coupled-cluster method using a self-consistent reaction field model for the environment. Further relevant details are given in Supporting Information Section S7, focusing on important issues such as the solvent model and the treatment of open-shell electronic structures. All possible linkage isomers of the CN ligands around the iron centres, spin states, and conformational isomers were considered. The only stable low-energy species predicted are the anticipated linkage isomers $[\{(NC)_5Fe^{II}\}(\mu\text{-}CN)\{Fe^{III}(CN)_5\}]^{6-}$ and $[\{(NC)_5Fe^{III}\}(\mu\text{-}CN)\{Fe^{II}(CN)_5\}]^{6-}$. Their ground states were predicted to arise from a Jahn-Teller distortion of a $^2E$ state, in agreement with the EPR data, but rendering them unsuitable for accurate computations using standard DFT methods. However, their energies were predicted to be similar, in qualitative agreement with the observed zero-phonon energy of just $h\nu_{00} = 0.33$ eV. The coupled-cluster SAC-CI method was used to perform quantitative calculations on these isomers and their IVCT transition, correctly predicting $[\{(NC)_5Fe^{II}\}(\mu\text{-}CN)\{Fe^{III}(CN)_5\}]^{6-}$ to be the ground state, with an IVCT band centred at $h\nu_v = 0.34$ eV and a transition moment $M = 1.8$ eÅ that is in reasonable agreement with observation (0.33 eV and 0.88 eÅ). This leads to:

- <u>Deduction 7:</u> Both CAM-B3LYP and coupled-cluster approaches predict detailed properties of the model compound and hence are expected to provide realistic descriptions of the chromophores in Prussian blue.

**Modelling the spectroscopic properties of dimer and tetramer models for Prussian blue.** Whilst the $[Fe_2(CN)_{11}]^{6-}$ dimeric mixed-valence ion provides several appealing features with regards to use as a Prussian blue model, the $Fe^{III}$ ion has a different spin state and coordination environment. Two unique atomic planes exist in the $\{Fe^{II}\}_3\{Fe^{III}\}_4(CN)_{18}(H_2O)_6$ Prussian blue framework giving rise to two types of "dimers", of form $[\{(NC)_5Fe^{II}\}(\mu\text{-}CN)\{Fe^{III}(NC)_3(H_2O)_2\}]^{4-}$ and $[\{(NC)_5Fe^{II}\}(\mu\text{-}CN)\{Fe^{III}(NC)_5\}]^{6-}$, of which the former is twice as prevalent (Fig. 5). Historically, exciton couplings within the crystal have assumed to be weak, facilitating the use of only dimer models in spectral interpretation (38, 40). Nevertheless, one of each type of dimer can be paired together into "tetramers" [{Fe$^{II}$}$_2${Fe$^{III}$}$_2$(CN)$_{18}$(H$_2$O)$_2$]$^{8-}$ (Fig. 5d) that depict a fundamental structural unit within Prussian blue, with differently oriented tetramers combining to make the crystal unit cell.

Here, TDDFT calculations have been performed on the model dimeric and tetrameric compounds based on the structural features identified in Fig. 5, as well as coupled-cluster calculations performed using EOM-CCSD. The intent of these calculations is to mimic conditions found in the crystal, achieved using external point charges to represent the external crystal field. It is not suitable to use a dielectric-response SCRF model, as was used to model [Fe$_2$(CN)$_{11}$]$^{6-}$ in aqueous solution, as the electrostatic potential so generated is opposite in nature to that coming from the surrounding crystal. Hence external point charges are fitted to reproduce the external field of the crystal, obtained using a Bader-charge representation of the periodic material, as described in Supporting Information Section S7. As this approach is crude and not performed self consistently, only qualitative predictions of IVCT transition energies are possible.

Calculated stick spectra for the model dimers and tetramer are reported in Fig. 6, obtained by applying a fixed and very narrow Gaussian spectral width to each vertical excitation energy and transition moment evaluated. CAM-B3LYP and EOM-CCSD calculations yield qualitatively similar results, whereas HSE06 results are starkly different. The HSE06 transition energies appear to be too low by 1 – 2 eV, and for the tetramer this method fails to correctly identify the ground state (as revealed by the predicted negative transition energy). This leads to

- Deduction 8: HSE06 appears unsuitable for the calculation of spectra involving charge transfer.

Both the CAM-B3LYP and EOM-CCSD calculations predict that [{(NC)$_5$Fe$^{II}$}(μ-CN){Fe$^{III}$(NC)$_3$(H$_2$O)$_2$}]$^{4-}$ absorbs 1.5 – 2 eV lower in energy than does [{(NC)$_5$Fe$^{II}$}(μ-CN){Fe$^{III}$(NC)$_5$}]$^{6-}$ and absorbs 2.4 – 11 times stronger. As a consequence, in the tetramer, the IVCT band for [{(NC)$_5$Fe$^{II}$}(μ-CN){Fe$^{III}$(NC)$_5$}]$^{6-}$ becomes lost amongst the overlapping MLCT and LMCT transitions, whereas the intense [{(NC)$_5$Fe$^{II}$}(μ-CN){Fe$^{III}$(NC)$_3$(H$_2$O)$_2$}]$^{4-}$ transition moves towards that observed in Prussian blue. The commonly used assumption that the IVCT band corresponds to the bandgap transition is not vindicated, with instead many forbidden transitions being predicted at energies below the allowed IVCT band (Fig. 6).

The CAM-B3LYP and EOM-CCSD calculated IVCT spectra show internal structure, the most distinct feature of which is that [{(NC)$_5$Fe$^{II}$}(μ-CN){Fe$^{III}$(NC)$_3$(H$_2$O)$_2$}]$^{4-}$ presents transitions of the same primary nature that are separated in energy based on the orientation of the Fe(III) receiving *d*-orbital with respect its water ligands. Transitions involving *d* orbitals in-plane with the oxygen atoms (Fig. 5a) appear at lower energy than those involving orbitals that are in-plane with κ-NC ligands (Fig 5b). In addition, the calculated spectra for the tetramer reflect the influence of weak exciton couplings. All of these results lead to

- <u>Deduction 9:</u> The predicted spectra of PB model compounds indicates that IVCT absorption can occur over a very wide energy range depending on Fe(III) ligation, with the lowest-energy transitions being forbidden, intense transitions arising from localised transitions within [{(NC)$_5$Fe$^{II}$}(μ-CN){Fe$^{III}$(NC)$_3$(H$_2$O)$_2$}]$^{4-}$ "dimers" split by linkage isomerism, and weak high-energy transitions associated with [{(NC)$_5$Fe$^{II}$}(μ-CN){Fe$^{III}$(NC)$_5$}]$^{6-}$ "dimers" becoming lost amidst other transitions.

**Identification of the primary chromophore responsible for IVCT absorption in Prussian blue.** The spectral properties of Prussian blue, using both the recently established {Fe$^{II}$}$_3${Fe$^{III}$}$_4$(CN)$_{18}$(H$_2$O)$_6$ framework (Fig. 1a) and the originally proposed K[Fe$^{II}$Fe$^{III}$(CN)$_6$] structure (Fig. 1b), both without water molecules of hydration, have been evaluated using the CAM-B3LYP and HSE06 density functionals; optimised coordinates are provided in Supporting Information Section S9. The results are given in Table 2 and directly parallel those obtained for the model dimers and tetramer (Deductions 8 – 9).

For the {Fe$^{II}$}$_3${Fe$^{III}$}$_4$(CN)$_{18}$(H$_2$O)$_6$ framework, CAM-B3LYP predicts a lowest vertical absorption energy of $h\nu_v$ = 1.62 eV, within the range of the values observed for soluble and insoluble Prussian blues (Fig. 2), whereas $h\nu_v$ is predicted to be 3.41 eV for K[Fe$^{II}$Fe$^{III}$(CN)$_6$] (see Deduction 9). This leads to

- <u>Deduction 10:</u> The primary chromophores responsible for the intense red IVCT transition in all Prussian blues are localised dimers of the form [{(NC)$_5$Fe$^{II}$}(μ-CN){Fe$^{III}$(NC)$_3$(H$_2$O)$_2$}].

HSE06 also predicts a large difference in the lowest excitation energy of these frameworks, but the energies are seriously underestimated owing to the asymptotic-potential error in HSE06 (73). As a result of this error, HSE06 predicts that the $S$ = 5/2 ground-state of Prussian blue (50) involves spin that is distributed beyond the Fe$^{3+}$ ions. This result parallels the failure of HSE06 to identify the ground state of the tetramer model compound (Deduction 8). Table 2 also indicates that HSE06 is not suitable for the evaluation of the exciton binding energies that are critical to spectral interpretation, as is known (73, 74).

CAM-B3LYP predicts an exciton binding energy for the lowest-energy transition in PB of 0.70 eV (Table 2). This is much less than analogous results obtained for the model dimer (EOM-CCSD 10 eV, CAM-B3LYP 5.2 – 5.4 eV) and tetramer (CAM-B3LYP 1.9 – 2.3 eV).

Also, as the lowest-energy transition was forbidden in all model compounds, it is unlikely that the CAM-B3LYP results presented in Table 2 correspond to the intense IVCT band. Hence, to more accurately model the observed spectra, all low-energy transitions need to be considered and their transition moments calculated. Ignoring exciton binding, expected to reduce transition energies by ca. 0.7 eV (to become more compatible with observations), and exciton couplings, preliminary spectra can be predicted, and CAM-B3LYP results are given in Supporting Information Section S7. These depict the onset of absorption at 2.4 eV into weakly allowed states, with multiple intense bands culminating at a maximum at 2.7 eV, reflecting the properties found for the model dimers and tetramer.

To include electron-hole interactions, BSE calculations were performed using the CAM-B3LYP orbitals as input. This is an ab initio approach but at the level applied only gives reasonable results if the initial orbitals are realistic (85). The observed spectrum is also influenced by the reorganisation energy associated with the change in geometry induced by the MLCT transition. To model this, the ground-state vibrational modes were calculated by CAM-B3LYP and Huang-Rhys factors evaluated (86). These were applied universally to broaden the BSE spectra. The resultant spectrum is compared to the observed spectra in Fig. 2b. The calculated spectrum shows enhanced structure and is too broad, but otherwise is in good agreement with the observed spectra. In particular, the enhanced structure parallels that observed in Fig. 4 in the MCD spectra. Paralleling Deduction 9,

- Deduction 11: The identification of two IVCT band components in the MCD spectra presented in Fig. 3 (Deduction 5) is attributed to a splitting, calculated to ca. 0.4 eV in magnitude, between the IVCT energies obtained when the receiving Fe(III) ion presents either water or κ-NC ligands in-plane with the $d$ orbital involved in the transition (Figs. 2 and 5).

**Conclusions**

A combination of absorption, MCD, and EPR measurements on a comprehensive range of Prussian blues, as well as a dimeric model ion, with DFT and ab initio modelling, identify the nature of the primary IVCT chromophores in one of the world's most significant dyes. These results establish paradigms applicable over a wide range of coordination polymers and MOFs.

Important results include the demonstration that no significant chromophores exist in the near-IR region, that apparent "solubility" and "insolubility" is controlled by crystal shape effects rather than crystal size or chemical variation, that light transmission occurs through only very small particles, that light scattering from particles larger than the wavelength of light is a key feature controlling the apparent spectrum in the critical bio-transparent window, that the primary chromophore is [{(NC)$_5$Fe$^{II}$}(μ-CN){Fe$^{III}$(NC)$_3$(H$_2$O)$_2$}], and comes in two variants depending on Fe(III) ligand isomerism, and that DFT functionals like HSE06 without long-range correction are inappropriate for the simulation of IVCT (and other) charge transfer spectra.

**Materials and methods**

The synthetic methods (2, 44, 47, 49, 51, 87) are described in detail in Supporting Information Section S1, with particle-sizing methods and results described in Supporting Information Section S2. Absorption spectroscopic measurements and additional results are described in Supporting Information Section S3, with MCD measurements (88) and the molecular modelling of their saturation data (89) described in Supporting Information Section S4. Continuous-wave EPR measurements (82) are described in Supporting Information Section S5, with pulsed X-band EPR measurements(83, 84) described in Supporting Information Section S6. The computational modelling (60, 73, 80, 90-96) involves the development of new strategies and is described in Supporting Information Section S7.

**Acknowledgements**

We thank the Australian Research Council for funding this work through grant DP190100074, as well as National Computational Infrastructure Australia (NCI) and the Shanghai University ICQMS Facility for computer time. P.S. gratefully acknowledges a scholarship and fees support from the UWA Research Training Program. We thank Prof. Deanna D'Alessandro (Sydney University) for helpful discussion and resources. We thank Microanalysis Australia for use the Malvern 2000 and Ms. Nimue Pendragon for help with these measurements. Mr Oscar Del Borello is thanked for technical assistance. Mr. Sebastian Rossi is thanked for assistance with the synthesis of the directly synthesised samples and room temperature spectroscopic measurements. The authors acknowledge the facilities, and the scientific and technical assistance of Microscopy Australia at the Centre for Microscopy, Characterisation & Analysis, The University of Western Australia, a facility funded by the University, State and Commonwealth Governments. We acknowledge the School of Molecular and Life Sciences at Curtin University for granting us access to the Malvern Zetasizer for DLS measurements.

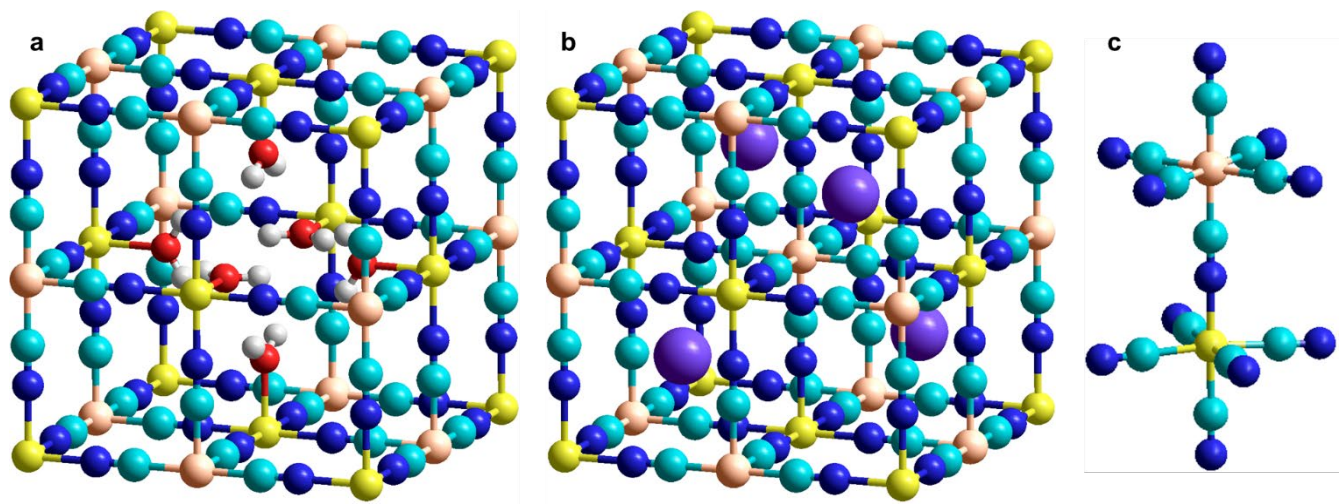

**Fig. 1.** Structures pertaining to Prussian blue dyes and their model ion. (a) The framework $\{Fe^{II}\}_3\{Fe^{III}\}_4(CN)_{18}(H_2O)_6$, now believed to be critical to all Prussian blue samples. (b) Translationally equivalent unit cell pertaining to $KFe^{II}Fe^{III}(CN)_6$, as originally suggested as the structure of soluble Prussian blue samples. (c) The synthesised model dimeric ion $[Fe_2(CN)_{11}]^{6-}$, identified as Jahn-Teller distorted $^2E$ $[\{(NC)_5Fe^{II}\}(\mu\text{-}CN)\{Fe^{III}(CN)_5\}]^{6-}$. Fe(III)- yellow, Fe(II-) orange, O- red, N- blue, C- cyan H- white, $K^+$- purple.

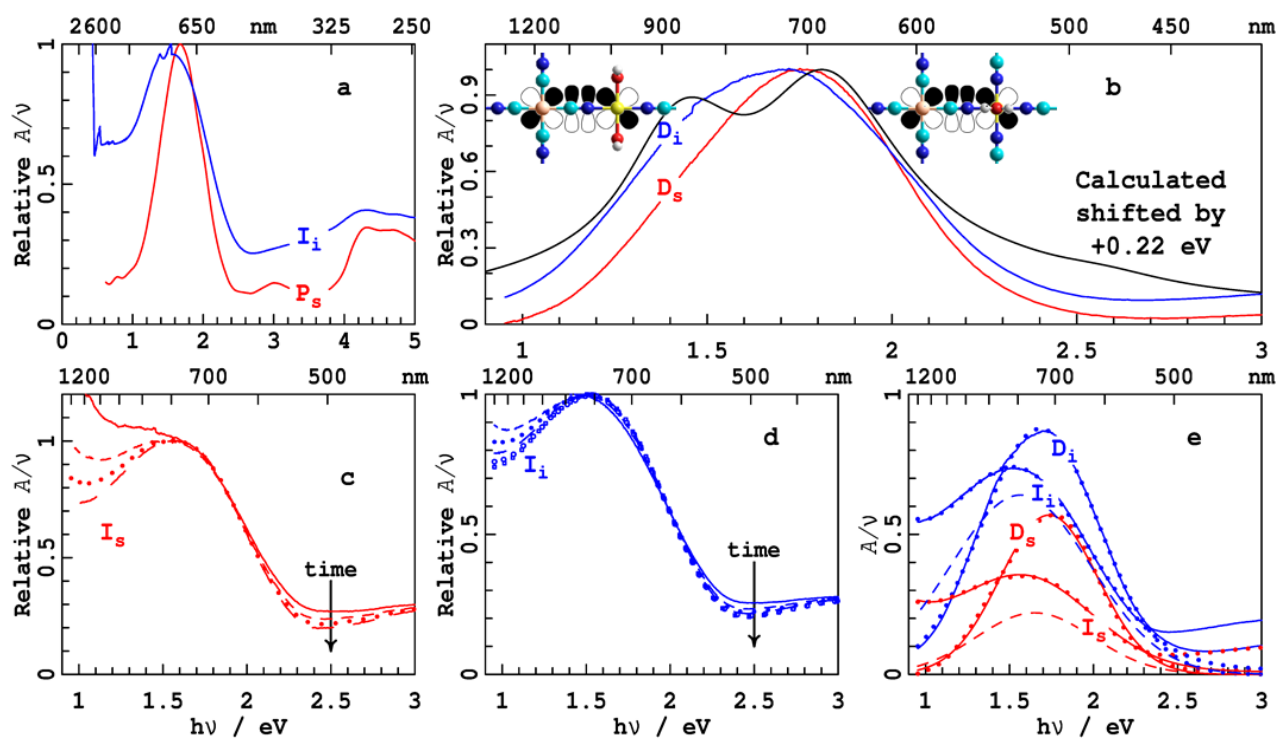

**Fig. 2.** IVCT spectra of Prussian blue. Red- soluble Prussian blues, blue- insoluble Prussian blues. (a) Comparison of Robin's narrow spectrum $P_s$ with a broad insoluble Prussian blue spectrum $I_i$. (b) Spectra of soluble $D_s$ and insoluble $D_i$ Prussian blue samples from direct syntheses compared with the CAM-B3LYP/BSE calculated spectrum (black), with the inserts depicting the chromophores responsible for the two anticipated spectral components (see Fig. 5). (c) Spectra of soluble Prussian blue $I_d$ recorded at time intervals of 7 min. (d) Spectra of insoluble Prussian blue $I_i$ recorded at time intervals of 7 min. (e) Comparison of the final spectra in the time sequences of $D_s$, $D_i$, $I_s$, and $I_i$ (see Supporting Information Section S3): solid- observed, dots- fit to Gaussian absorption plus power-law scattering for $I_s$ and $I_i$, dashed- absorption component thereof.

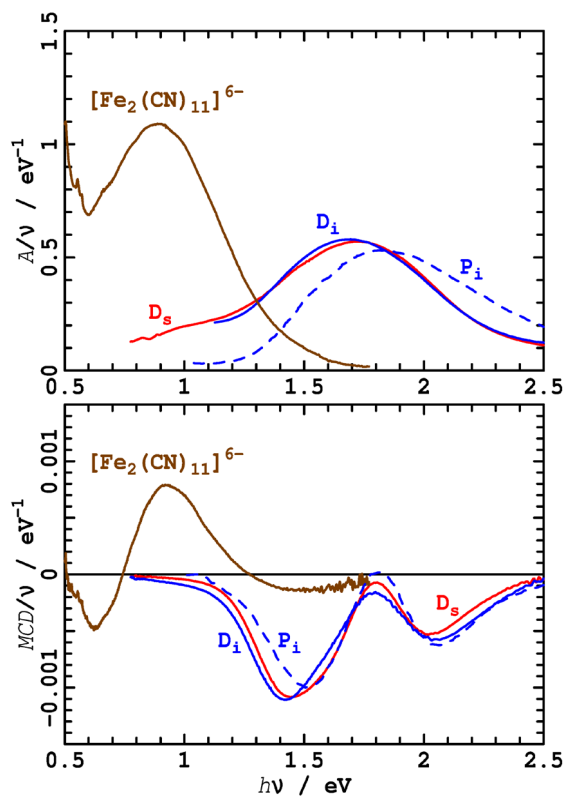

**Fig. 3.** Absorption and MCD bandshapes for soluble Prussian blue $D_s$ and insoluble Prussian blues $D_i$ and $P_i$ and the $[Fe_2(CN)_{11}]^{6-}$ ion. $D_s$ and $D_i$ were measured in a 50% glycerol/water matrix at 1 T field strength and 1.8 K, whilst $P_i$ was measured in a PVA film in absorption at 10 K 5 T and in MCD at 1.8 K (as recorded in 1976). The $[Fe_2(CN)_{11}]^{6-}$ ion was measured at 5 T field strength and 1.8 K.

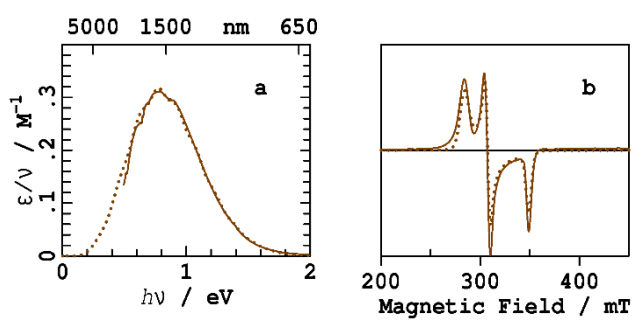

**Fig. 4.** Solution spectra of $[Fe_2(CN)_{11}]^{6-}$. (a) In aqueous solution: solid- IVCT absorption spectrum, dots- its fit to a Huang-Rhys model ($h\nu_{00} = 0.33$ eV, $h\nu_v = 0.86$ eV, $v = 1200$ cm$^{-1}$, $S = 3.56$, $M = 0.876$ eÅ, Gaussian broadened with resolution 1190 cm$^{-1}$). (b) solid- X-band CW frozen-solution spectrum at 10 K, dots- EasySpin *pepper* simulation (82) (see SI Section S5.

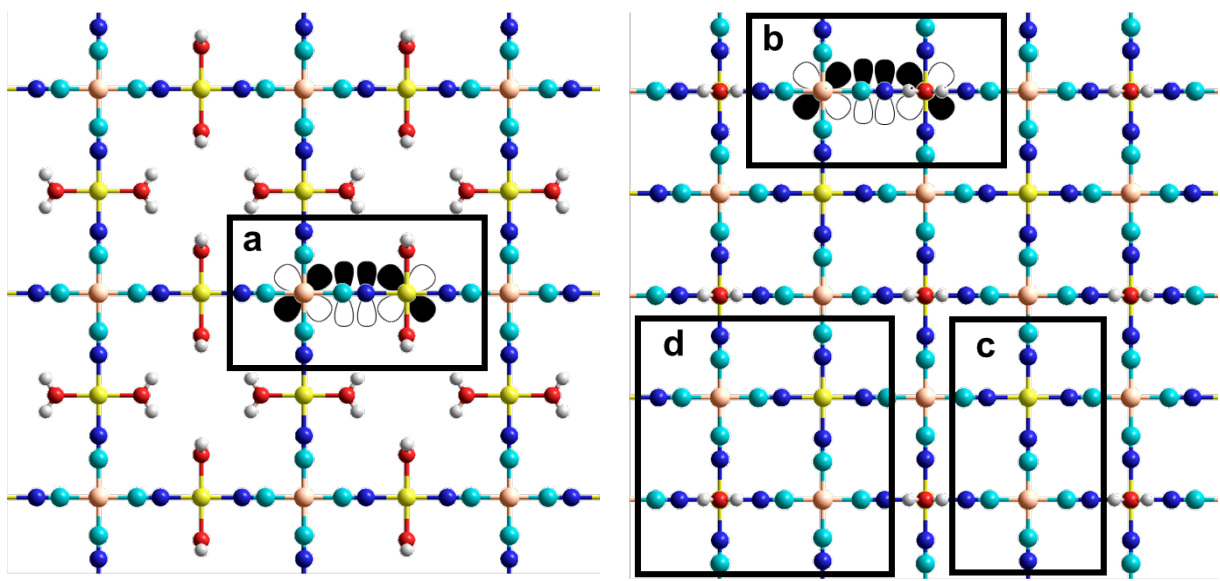

**Fig. 5.** Substructures within the two planes comprising the primary Prussian blue framework $\{Fe^{II}\}_3\{Fe^{III}\}_4(CN)_{18}(H_2O)_6$. (a) and (b) highlight "dimers" of the form $[\{(NC)_5Fe^{II}\}(\mu\text{-}CN)\{Fe^{III}(NC)_3(H_2O)_2\}]^{4-}$, with the shown $d$ and $p$ orbitals facilitating in-plane IVCT transitions in which the receiving $Fe^{III}$ orbital lies (a) in-plane with an attached oxygen ligand and (b) in-plane with an attached NC ligand. (c) highlights a "dimer" of the form $[\{(NC)_5Fe^{II}\}(\mu\text{-}CN)\{Fe^{III}(NC)_5\}]^{6-}$. (d) highlights a "tetramer" $[\{Fe^{II}\}_2\{Fe^{III}\}_2(CN)_{18}(H_2O)_2]^{8-}$ comprised of neighbouring $[\{(NC)_5Fe^{II}\}(\mu\text{-}CN)\{Fe^{III}(NC)_3(H_2O)_2\}]^{4-}$ and $[\{(NC)_5Fe^{II}\}(\mu\text{-}CN)\{Fe^{III}(NC)_5\}]^{6-}$ dimers. Fe(III)- yellow, Fe(II)- orange, O- red, N- blue, C- cyan, H-white; the dimer scenarios (a), (b), and (c) are equally prevalent in Prussian blue.

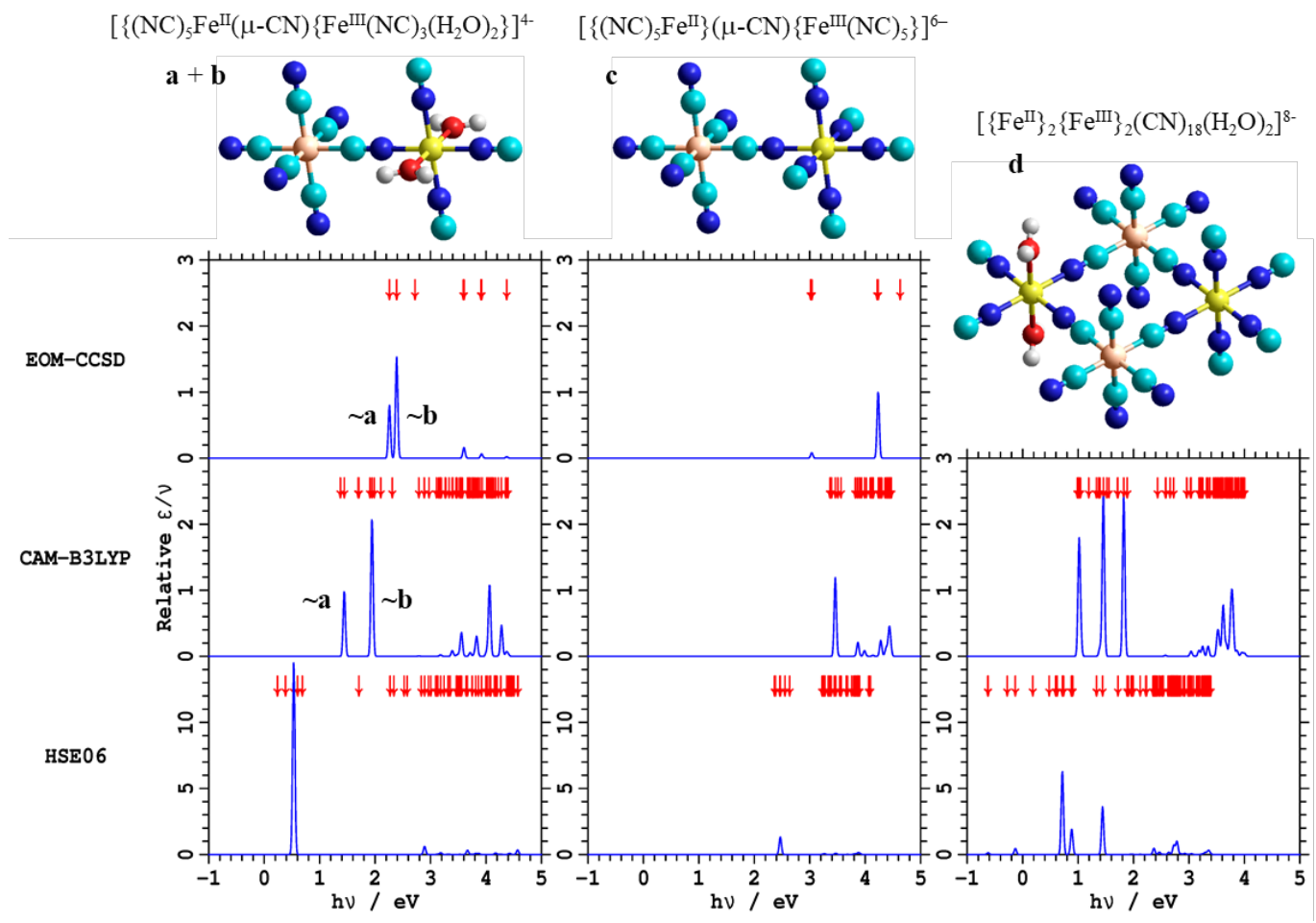

**Fig. 6.** Calculated stick spectra for Prussian blue model ions featuring Fig. 5 elements **a** – **d**. These display the calculated vertical transition energies and relative intensities, using a constant narrow Gaussian bandwidth of 10 meV. Arrows indicate calculated vertical transition energies (only low-energy states of each symmetry are shown for EOM-CCSD). Fe(III)- yellow, Fe(II)- orange, O- red, N- blue, C- cyan, H-white

**Table 1.** Properties and synthesis conditions of the Prussian blues from previous reports (P) and new direct (D) or indirect (I) syntheses.

|  | synthesis | | | key properties | | | IVCT band | |
|---|---|---|---|---|---|---|---|---|
|  | type | Env. | Time | type[c] | % K | size / μm | $h\nu_v$ / eV | FWHM / eV |
| $P_s$[a] | direct (38), ferric + ferrocyanide | HClO$_4$ |  | soluble |  | ≪ 0.2 | 1.68 | 0.80 |
| $P_i$[b] | direct (49), ferric + ferrocyanide | HCl | 8 weeks | insoluble | 0.9 | 150[d] | 1.84 | 0.85 |
| $D_s$ | direct (2), ferric + ferrocyanide in ratio 2:3 | H$_2$O | minutes | soluble | 11 | 70[d] | 1.74[e] | 0.68[e] |
| $D_i$ | direct (2), ferric + ferrocyanide in ratio 5:6 | H$_2$O | minutes | insoluble | 7 | 80[d] | 1.68[f] | 0.83[f] |
| $I_s$ | indirect (44), ferrous + ferrocyanide in ratio 1:1 | HCl | 2 weeks | soluble | 2.7 | 30[d] | 1.60 | 0.92 |
| $I_i$ | indirect (49), ferrous + ferrocyanide in ratio 3:1 | HCl | 8 weeks | insoluble | 1.4 | 30[d] | 1.57 | 0.98 |

a: previously reported spectrum (38) following synthesis, in boiling perchloric acid, designed to create extremely small particles.
b: previously reported spectrum (49).
c: determined based on whether the majority of material in suspension precipitated within a day, see Supporting Information Video.
d: most common size, with particles as small as 1 – 3 μm also apparent, see Supporting Information Figs. S1-2; these do not contribute to the spectra, however, as the total path length through all crystals in the light beam is estimated to be ca. 0.18 μm.
e: in 50% glycerol/water at 5 K, $h\nu_v$ = 1.71 eV, FMHM = 0.86 eV.
f: in 50% glycerol/water at 5 K, $h\nu_v$ = 1.70 eV, FMHM = 0.85 eV.

**Table 2.** Calculated bandgap, exciton binding energy, and lowest vertical transition energies $h\nu_v$ for Prussian blue framework $Fe_3^{II}Fe_4^{III}(CN^-)_{18}(H_2O)_6$ (Fig. 1a), as well as that for the originally proposed $Fe^{II}Fe^{III}(CN^-)_6 \cdot K^+$ framework.

| Functional | Structure | Band gap / eV | Exciton binding energy / eV | $h\nu_v$ / eV |
|---|---|---|---|---|
| CAM-B3LYP | {Fe$^{II}$}$_3$ {Fe$^{III}$}$_4$(CN)$_{18}$(H$_2$O)$_6$ | 2.33 | 0.70 | 1.62 |
|  | KFe$^{II}$Fe$^{III}$(CN)$_6$ | 3.96 | 0.55 | 3.41[a] |
| HSE06 | {Fe$^{II}$}$_3$ {Fe$^{III}$}$_4$(CN)$_{18}$(H$_2$O)$_6$ | 0.70[b] |  |  |
|  | KFe$^{II}$Fe$^{III}$(CN)$_6$ | 2.26[c] | 0.03 | 2.23 |

a: 3.8 eV as calculated by multi-reference DFT-based methods (56).
b: this is for a ferromagnetic state with Fe(III) high spin and Fe(II) low spin, as observed, however HSE06 predicts that the ground state is actually one in which the spin is distributed over both ions.
c: other calculations (58) give 2.31 eV.

# Identification of the Chromophore in Prussian blue

*Li Musen,[1,2]\* Robin Purchase[3]\* Parvin Safari[4]\*, Martyna Judd,[3]\* Emily C. Barker,[4,5] Jeffrey R. Reimers,[1,2]\* Nicholas Cox,[3] Paul J. Low,[4] and Elmars Krausz[3]*

1 International Centre for Quantum and Molecular Structures and Department of Physics, Shanghai University, Shanghai 200444, China.
2 University of Technology Sydney, School of Mathematical and Physical Sciences, Ultimo, NSW 2007, Australia.
3 Research School of Chemistry, Australian National University, Canberra, ACT 2601 Australia.
4 School of Molecular Sciences, University of Western Australia, Perth W6009, Australia
5 Centre for Microscopy, Characterisation and Analysis, University of Western Australia, Perth, WA 6009, Australia

Email:    jeffrey.reimers@uts.edu.au,   robin.purchase@anu.edu.au,   paul.low@uwa.edu.au, nick.cox@anu.edu.au, elmars.krausz@anu.edu.au

## Contents

# S1. Synthesis and basic characterisation.

**General Information:**

The reagents $Fe_2(SO_4)_3.9H_2O$, $FeCl_2.4H_2O$, $K_3[Fe(CN)_6]$, and $K_4[Fe(CN)_6].3H_2O$ were purchased as reagent grade and used without further purification. The complex $Na_3[Fe(CN)_5.NH_3].H_2O$ was prepared according established procedure(1).

Prussian blues are typically prepared by 'direct' or 'indirect' methods. The direct method involves combination of aqueous solution of ferrous ($Fe^{II}$) and ferricyanide ($[Fe^{III}(CN)_6]^{3-}$) ions, or ferric ($Fe^{III}$) and ferrocyanide ($[Fe^{II}(CN)_6]^{4-}$ ions. The 'indirect' methods use both reagents in the $Fe^{II}$ state (i.e. ferrous ions and ferrocyanide) giving initially homovalent $Fe^{II/II}$ ferrous ferrocyanide framework material, also known as Berlin white, $Fe^{II}_2[Fe^{II}(CN)_6]$. Berlin white is then allowed to oxidise to give mixed-valence $Fe^{II/III}$ Prussian blue. The physical morphology and particle size distribution is highly sensitive to the preparation method, including details such the aging time employed during the formation of Berlin white and nature of the oxidising agent employed in the indirect method of synthesis or Ostwald ripening of the initially precipitated Prussian blue in its mother liquors. To overcome some of these variations, Ludi and colleagues modified the indirect route to prepare larger single crystals of Prussian blues from solutions of ferrous ions and ferrocyanide in concentrated HCl. Slow vapour diffusion of water and aerial oxidation over several weeks results in the formation of 'well-crystallised' Prussian blues(2). This is the indirect procedure adopted herein.

**a. Indirect methods**

*'Well-crystallised' Insoluble Prussian blue* $I_i$

The 'well-crystallised' sample of insoluble Prussian blue was prepared following an established method(2). Samples of $FeCl_2.4H_2O$ (746 mg, 3.75 mmol) and $K_4[Fe(CN)_6].3H_2O$ (528 mg, 1.25 mmol) were separately dissolved in minimum volumes of deionised water (15 ml). These solutions were then slowly and sequentially added to a relatively large volume of concentrated HCl solution (32%, 285 ml). The beaker containing the reaction mixture was placed in a desiccator containing deionised water (300 ml) in the style of a bain marie; the desiccator is employed as a simple reaction chamber, and is used free of any desiccant (!) . The stopcock of the desiccator was open to enable the diffusion of air and evaporation and the assembly left without disruption. After 4 weeks, dark blue, cubic-shaped Prussian blue crystals were collected by filtration, washed with deionised water carefully and dried under vacuum at 40 °C for one day. Analytical results from ICP-OES were in good agreement with other reports(3): Fe 33.1%, K 1.5 %.

*'Well-crystallised' Soluble Prussian blue* $I_s$

The 'well-crystallised' sample of soluble Prussian Blue was prepared following an established method(4). Samples of $FeCl_2.4H_2O$ (596 mg, 3.00 mmol) and $K_4[Fe(CN)_6].3H_2O$ (1267 mg, 3.00 mmol) were separately dissolved in minimum volumes of deionised water (15 ml), before being added slowly and sequentially to concentrated HCl solution (32%, 270 ml). The beaker containing this solution was placed in a desiccator containing deionised water (300 ml) as described above. The stopcock of the desiccator was left open to enable the slow evaporation of the solutions and ingress of air, and the solution left without disruption. After 2 days, a precipitate of Berlin white was observed, which evolved over time until after 2 months, "soluble" Prussian blue crystals were formed. These were collected by filtration, washed with deionised water carefully and dried under vacuum at 40 °C for one day. Analytical results from ICP-OES were in good agreement with previous reports(4): Fe 34.3%, K 3.7 %.

### b. Direct methods

*Sample* $D_s$

Following Cheng et al.(5), solutions of $Fe_2(SO_4)_3 \cdot 9H_2O$ (1.208 g, 2.15 mmol) in deionised water (45 ml) and $K_4[Fe(CN)_6] \cdot 3H_2O$ (1.381 g, 3.27 mmol) in deionised water (20 ml) were prepared separately before being rapidly mixed in an open vessel. To avoid loss of finely divided material, the resulting suspension of Prussian blue was used directly in analyses.

*Sample* $D_i$

Following Cheng et al.(5), solutions of $Fe_2(SO_4)_3 \cdot 9H_2O$ (1.499 g, 2.67 mmol) in deionised water (45 ml) and $K_4[Fe(CN)_6] \cdot 3H_2O$ (1.381 g, 3.27 mmol) in deionised water (20 ml) were prepared separately before being rapidly mixed in an open vessel. To avoid loss of finely divided material, the resulting suspension of Prussian blue was used directly in analyses.

### c. Model ion $K_xNa_y[Fe_2(CN)_{11}]^{6-}$

The model mixed-valence ion $[Fe_2(CN)_{11}]^{6-}$ was synthesised by a minor modification of the literature method(6), and isolated as a mixed $K^+/Na^+$ salt. The compounds $Na_3[Fe(CN)_5 \cdot NH_3] \cdot H_2O$ (100 mg, 0.307 mmol) and $K_3[Fe(CN)_6]$ (101 mg, 0.307 mmol) were separately dissolved in small volumes of deionised water (1.2 ml). Then, the $Na_3[Fe(CN)_5 \cdot NH_3] \cdot H_2O$ solution was dropwise added to $K_3[Fe(CN)_6]$ solution. The flask was wrapped with aluminium foil and the bright yellow solution was stirred overnight. The product was precipitated out as a pale-yellow solid by adding methanol. The solid was collected by filtration, washed with methanol and dried under vacuum for 10 h (85 mg). ICP-OES results: Fe 15.3%, K 25.4%, Na 4.8% corresponding to $K_{4.7}Na_{1.3}[Fe_2(CN)_{11}]$.

## S2. Particle sizing

### a. Particle Sizing by Laser Diffraction (PSLD)

The size distributions of the different samples of Prussian Blue were measured using with a Malvern 2000 Particle Size Analyzer by laser diffraction. This instrument measures the particle size between 10 nm and 2 mm by the diffraction of a blue laser from incident particles. To reduce aggregation the sample solution is constantly pumped through the cell during analysis, with a dispersant added to the solvent (see below). The instrument used is calibrated monthly using a Malvern Panalytical QAS 4002 standard. Measurements were conducted according to ISO 13320:2020 *Particle size analysis – Laser diffraction methods* with the dried sub-samples dispersed in a solvent of 0.01% v/v sodium hexametaphosphate in deionised water. Sample preparation was performed according to ISO 14887 *Sample preparation – Dispersing procedures for dry powders in liquids*. The larger end of the size distribution would again not have been reliable due to the strong blue colour of the sample; however, the powders were seen to have a maximum diameter much smaller than the upper limit of the instrument measurement range. Size quality reports, standard deviation of measurements and fit reports were again considered to ensure the reproducibility and statistical significance of the measurement and reliability of the conclusions drawn from these data. To limit aggregation in solution and precipitation, solutions for PSLD were prepared in deionised water containing 0.01% v/v sodium hexametaphosphate. The resultant particle-size distributions are shown in Fig. S1.

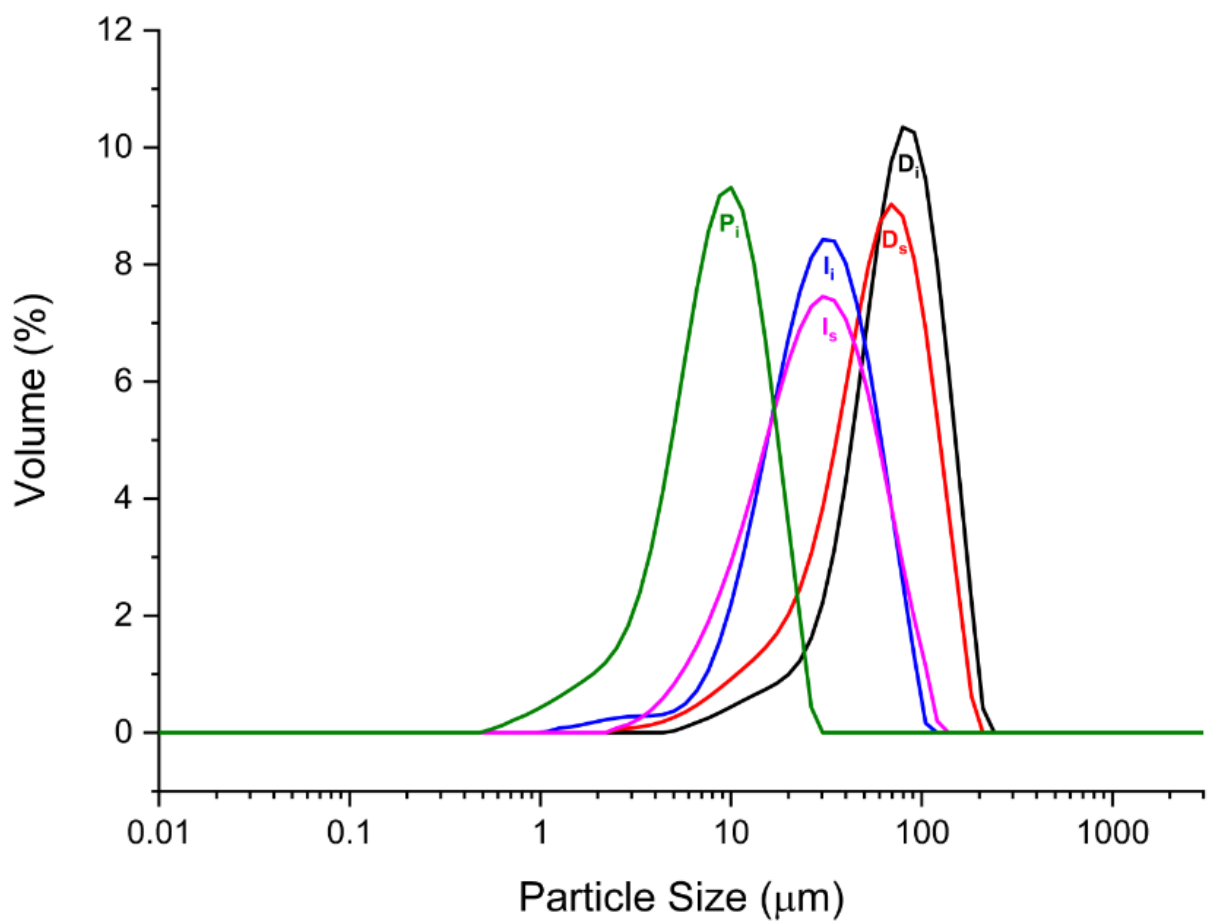

**Fig. S1.** Particle sizing by PSLD for samples $P_i$, $D_s$, $D_i$, $I_s$, and $I_i$.

**b. Scanning Electron Microscopy (SEM)**

Sub-samples of the powders were mounted on carbon conductive adhesive tabs on pin stubs before a platinum coating was applied (approximately 3 nm thickness). Secondary images of the powders were captured using a Zeiss 1555 VP-FESEM to analyse size and shape of the particles and aggregates. Data are summarised in Table S1, and representative micrographs, listing observation conditions, are presented in Fig. S2a-e. A diverse range of particle sizes is present in the images of samples $D_s$ and $D_i$ prepared by direct methods, with the anhedral morphology of particles being consistent with their formation through aggregation. In contrast, the samples $I_s$ and $I_i$ formed indirectly were comprised of a more uniform euhedral morphology, producing cubic crystals with more uniform dimensions. The $P_i$ sample presented as a mixture of particles with these two broad morphologies.

**Table S1.** Summary of particle sizes and morphologies determined from SEM imaging.[a]

| Sample | Size range (µm) | Morphology/Shape |
|---|---|---|
| $D_s$ | 5 - 70 | Irregular, higher aspect ratio |
| $I_s$ | 5 - 20 | Cubic with fractures |
| $D_i$ | 20 - 100 | Irregular, higher aspect ratio |
| $I_i$ | 5 - 30 | Cubic with fractures |
| $P_i$ | 2 - 15 | Mixture of cubic and irregular |

[a] The larger end of the size range is not imaged in the SEM results, likely a consequence of sampling difficulties arising from the large particle size distribution. Due to such sampling difficulties, SEM is not recommended as a technique for determining particle size distribution, but the range of particle sizes observed supports the distributions determined from other methods. The SEM does distinctly show the difference in morphology between the **D** and **I** samples with the $P_i$ sample being a mixture of the two.

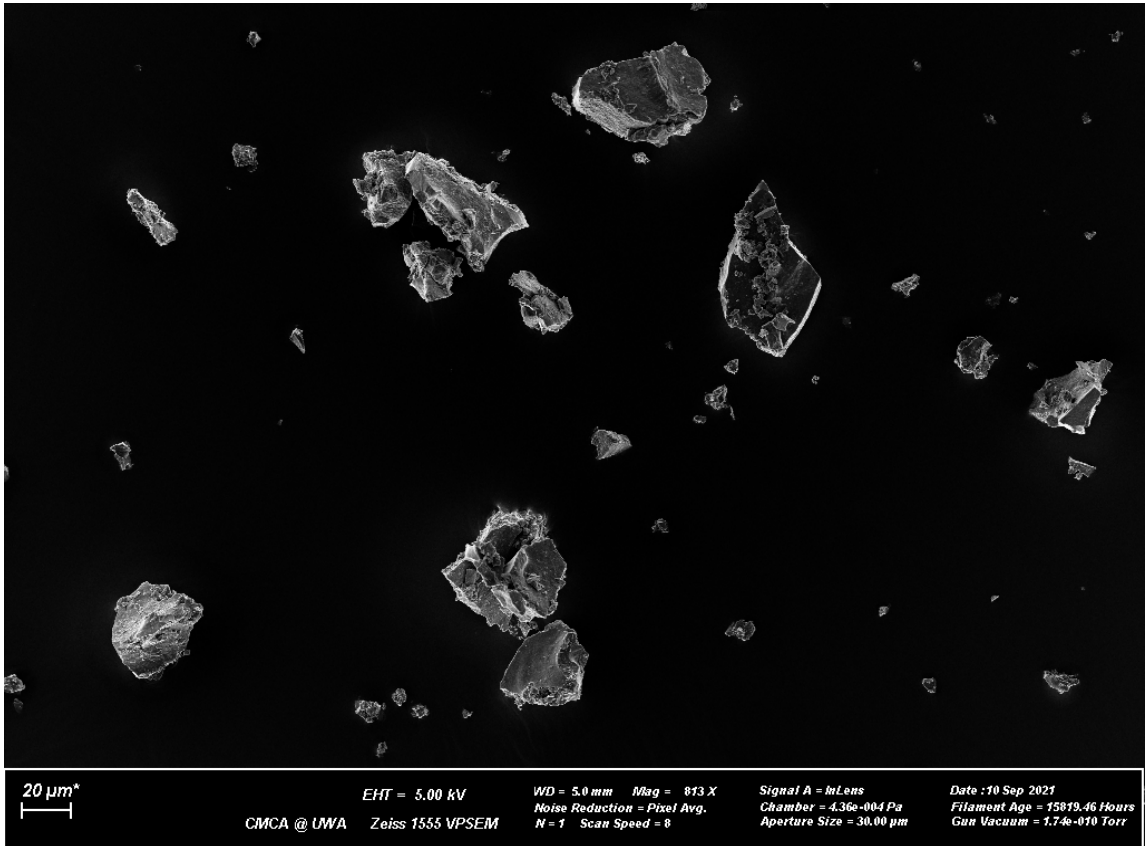

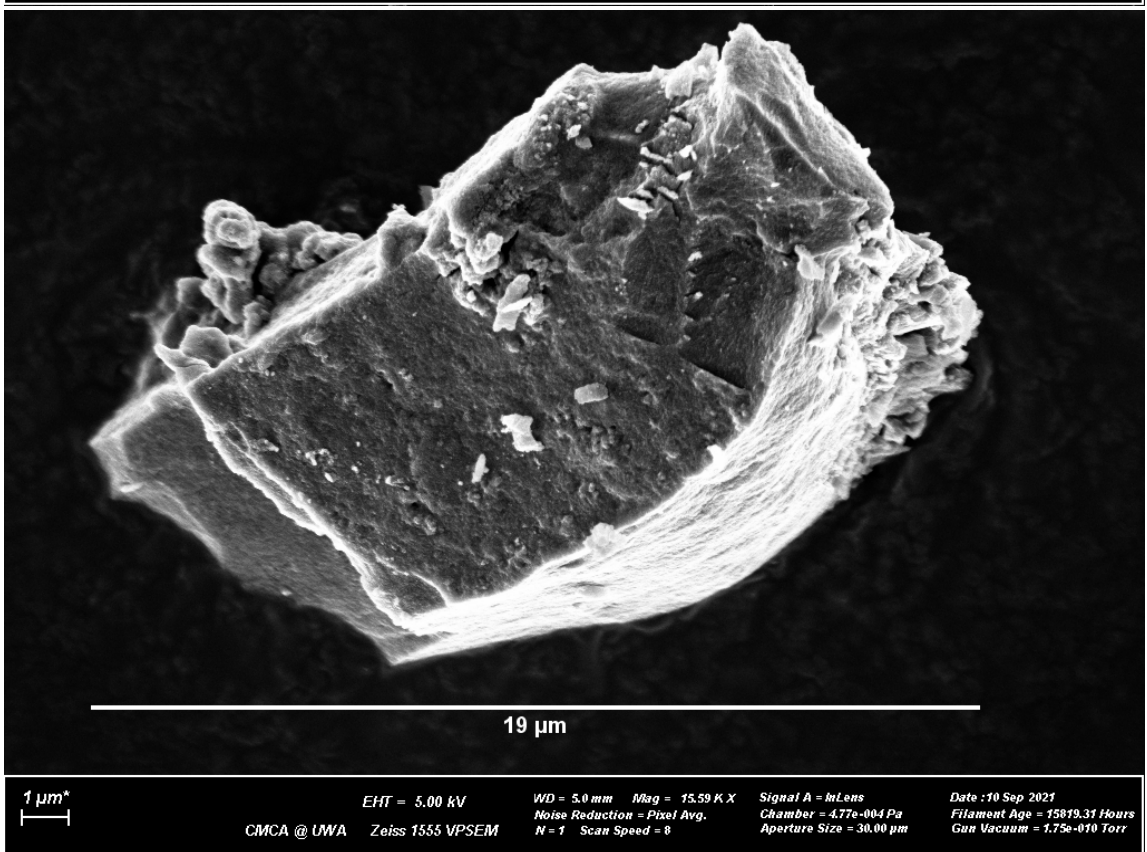

**Fig. S2a.** SEM images of $D_s$.

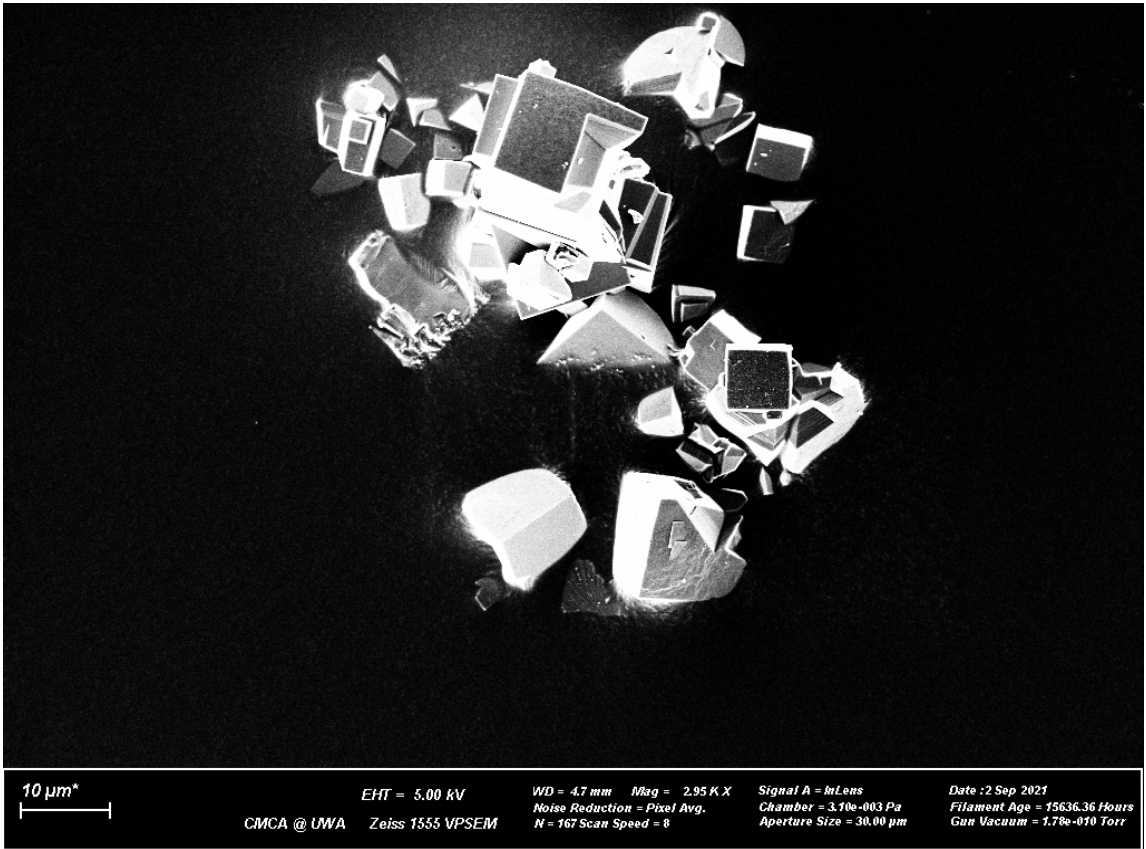

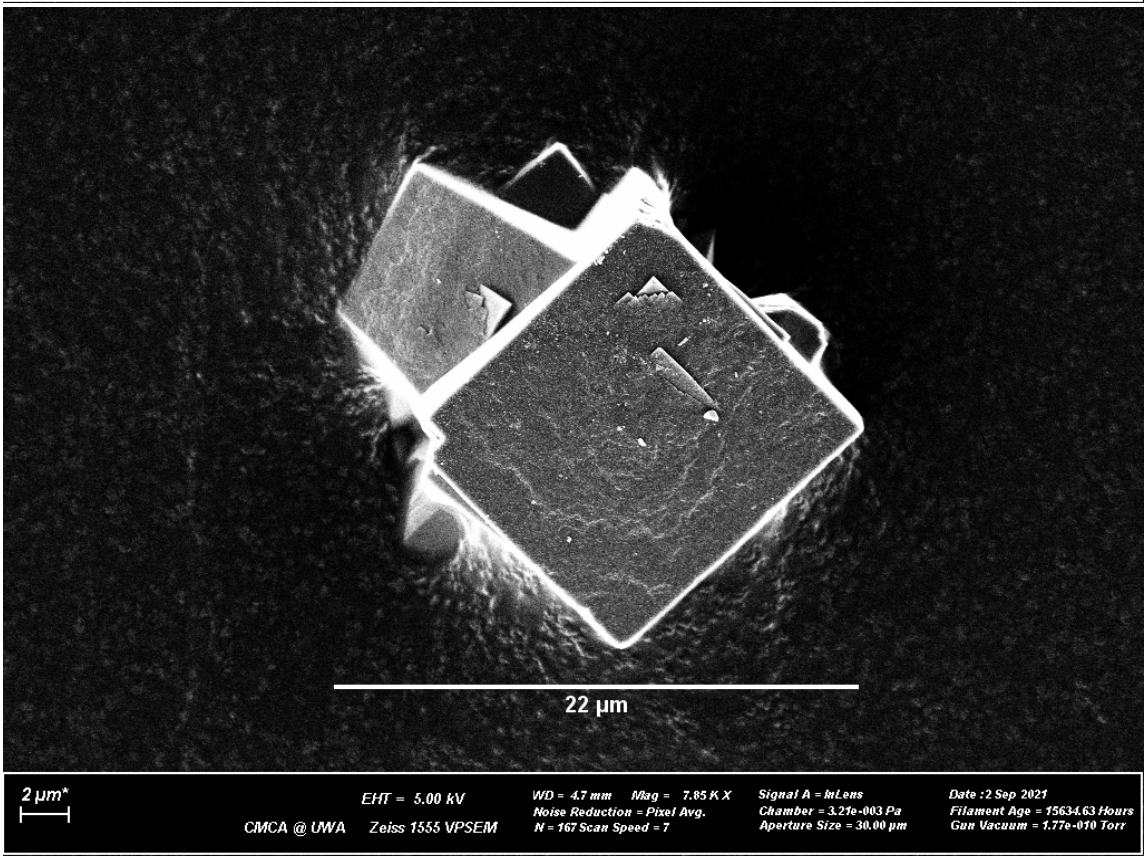

**Fig. S2b.** SEM images of **I**$_s$.

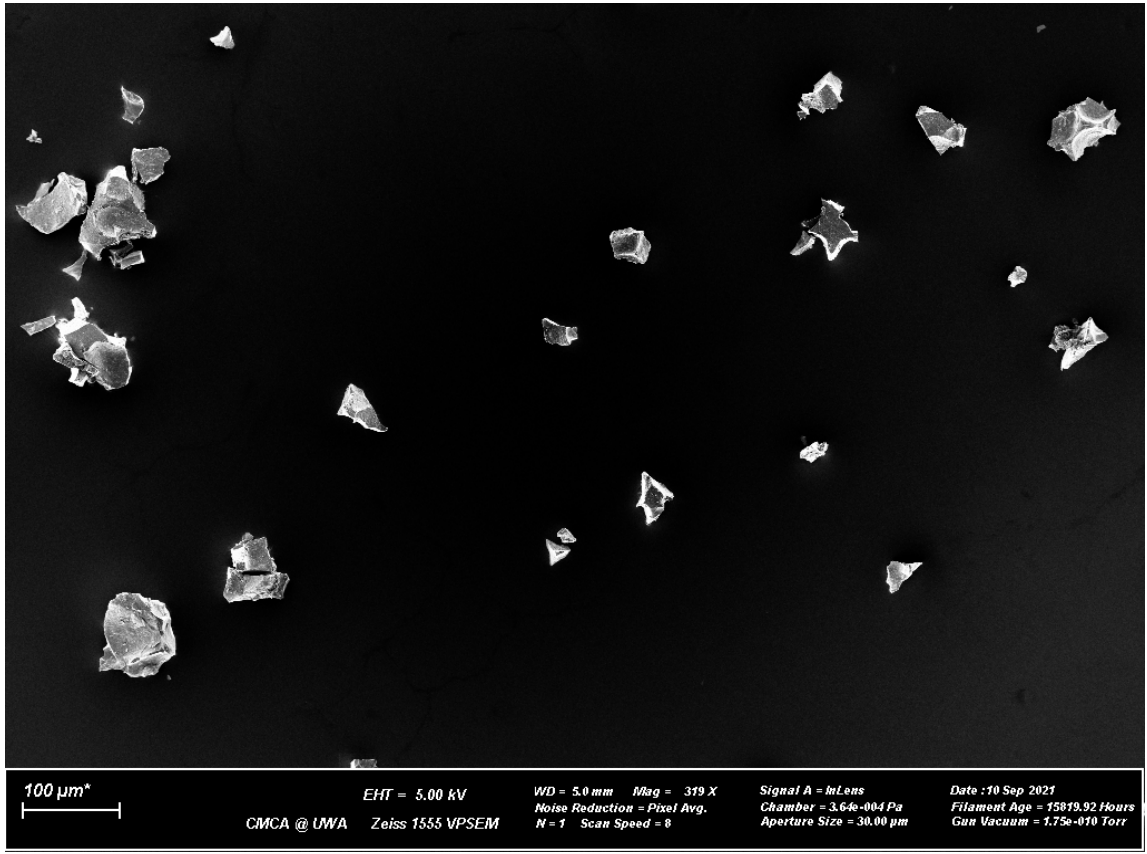
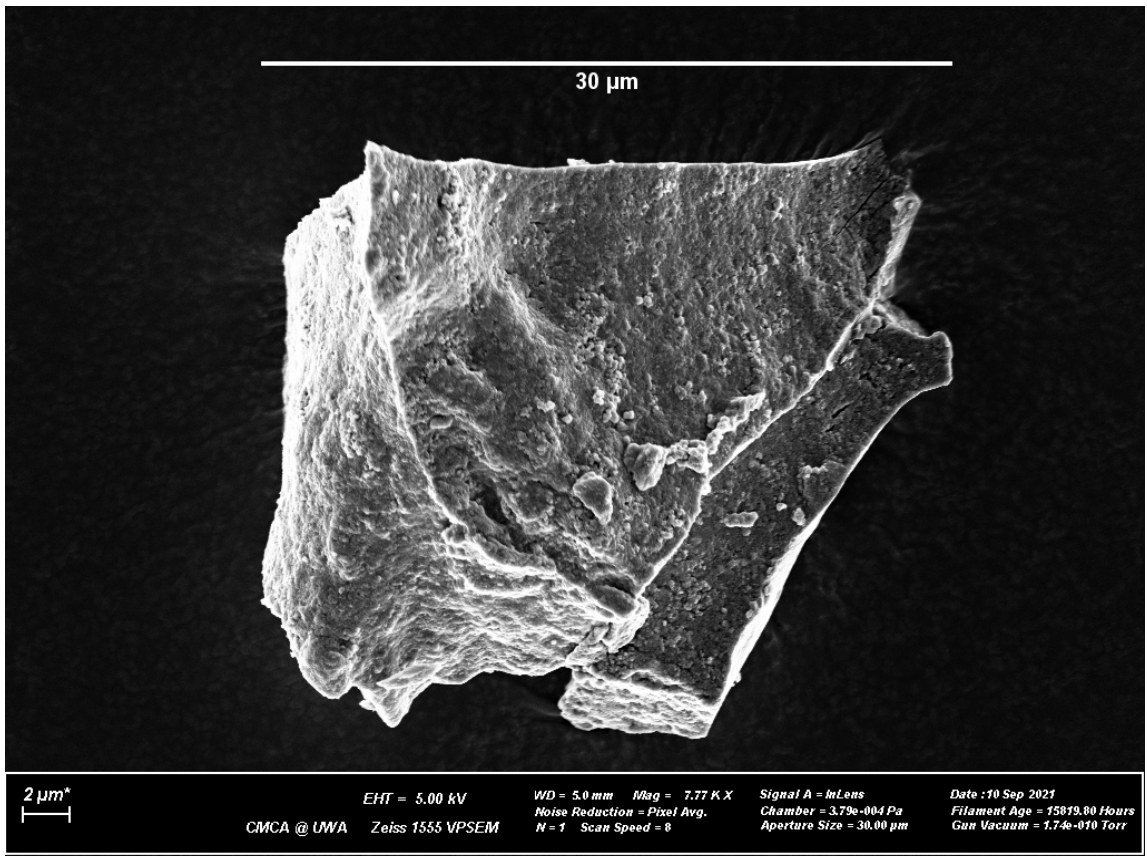

**Fig. S2c.** SEM images of **D**$_i$.

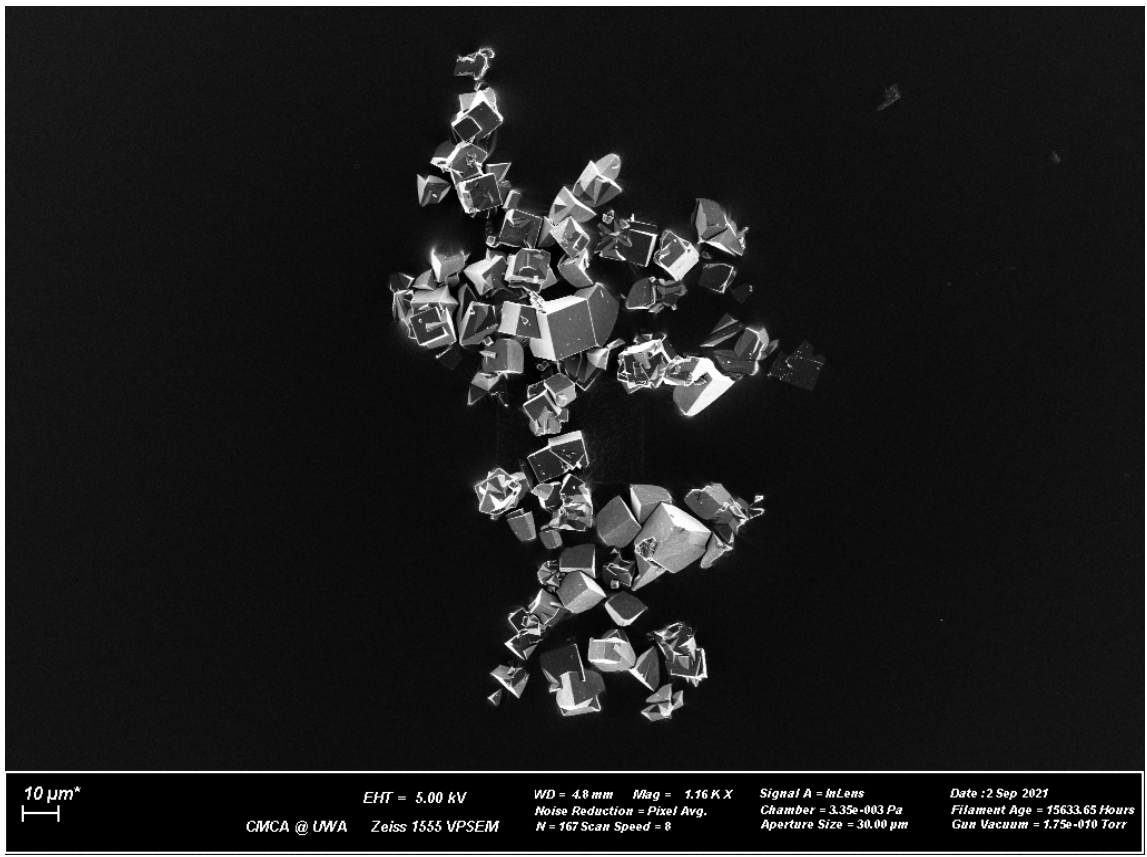

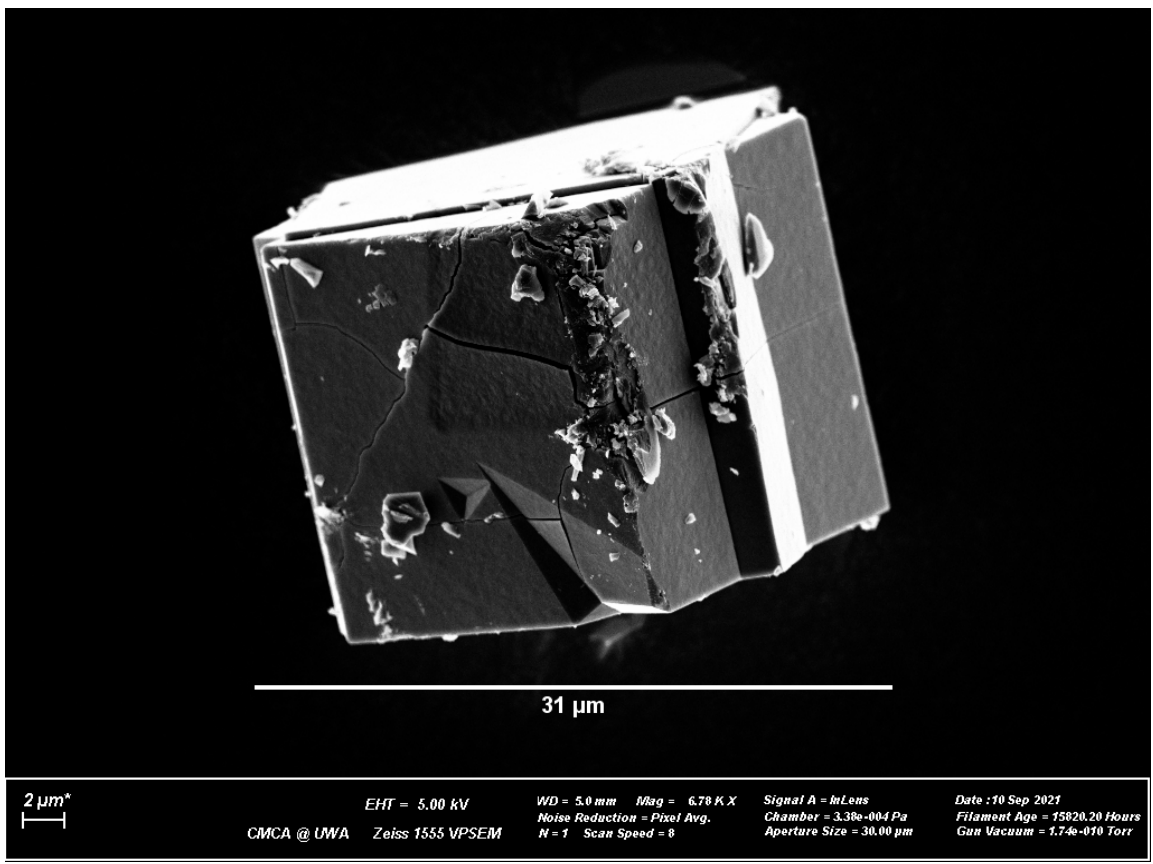

**Fig. S2d.** SEM images of **I**$_i$.

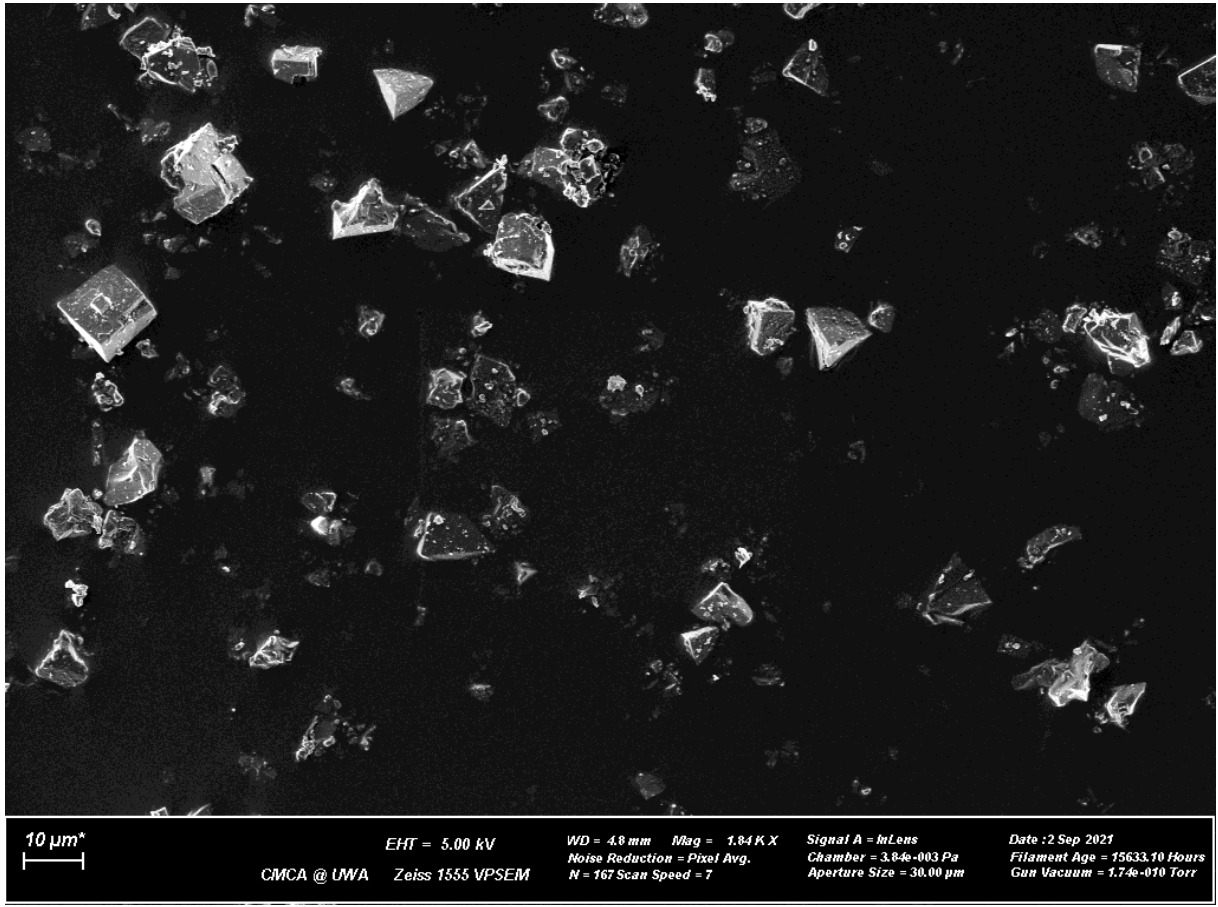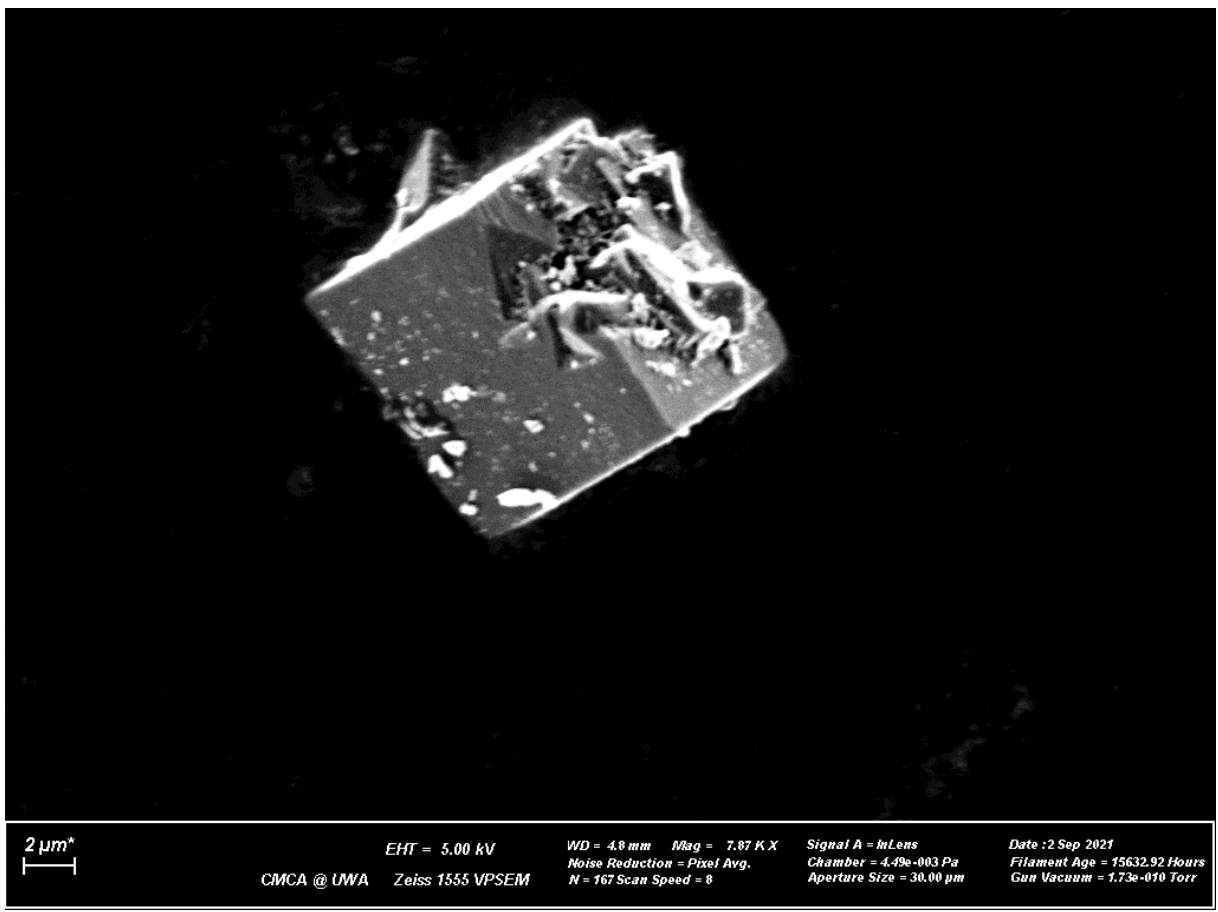

S11

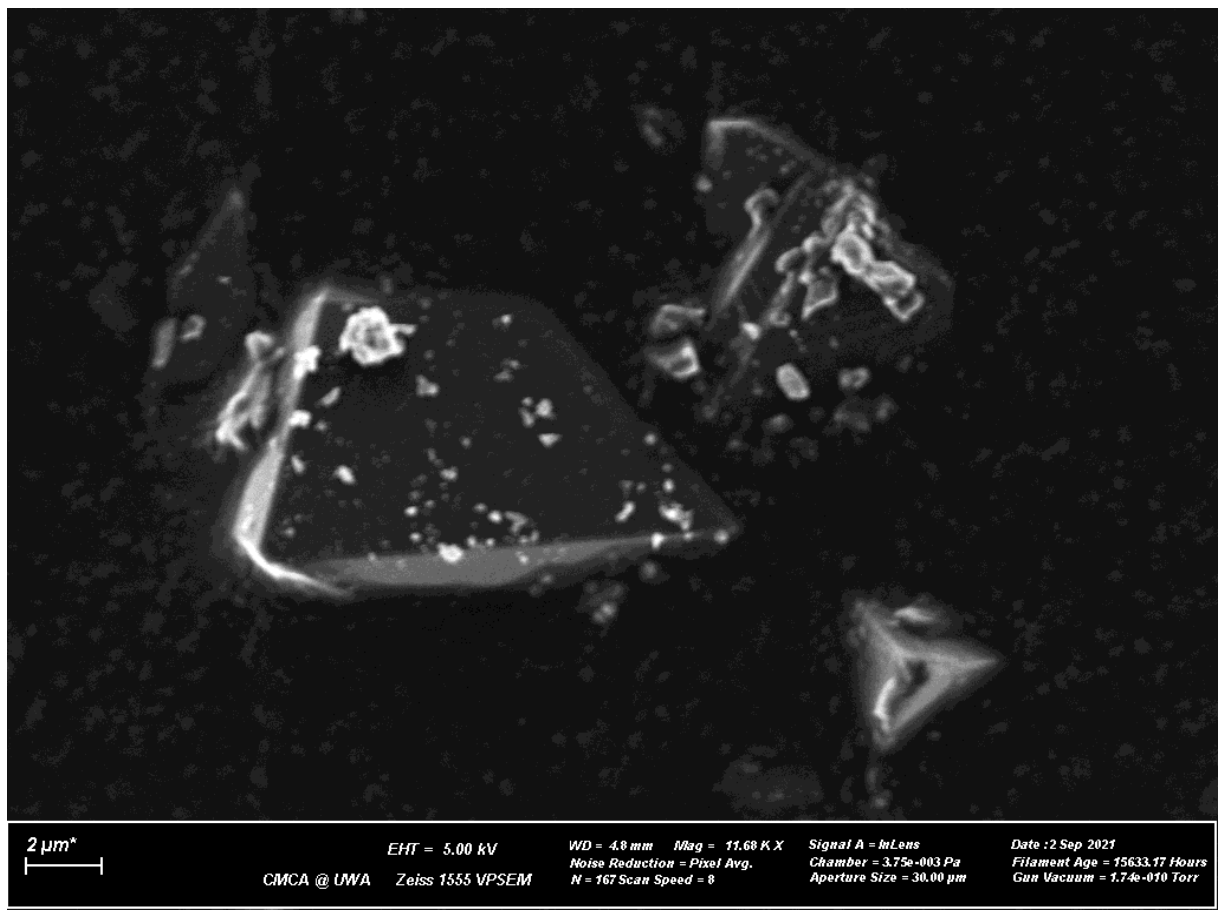

**Fig. S2e.** SEM images of $P_i$.

## S3. Absorption measurements.

### a. Prussian blue

For the IVCT spectrum of $I_i$ shown in Fig. 2a, the samples were ground and dispersed in Nujol and they were sandwiched between NaCl plates for the measurement of UV-vis-NIR spectra. The spectrum was recorded on a Cary 5000 spectrometer. The sharp onset of absorption at low energy in Fig. 2a arises from purely vibrational excitation.

The IVCT spectra of $D_s$, $D_i$, $I_s$, and $I_i$ shown in Figs. 2b-e were measured with a Cary 17 spectrometer in solution cells of length 1 cm, using a large-area light beam of order cm$^2$. Successive measurements were recorded 7 minutes apart. The IVCT band was fitted using a Gaussian function to describe the spectral bandshape $A(\nu)/\nu$, and a power-law expansion used to describe the wavelength dependence of the scattering off Prussian-blue particles using the expression

$$A(\nu) = B\nu^{-n} + A\nu \exp\frac{-(\nu - \nu_m)^2}{2\sigma^2} .$$

Here, $\nu_m$ is the centre of the absorption band, which is typically 0.05 – 0.2 eV less than the apparent band centre obtained by simply inspecting the raw absorption spectrum $A(\nu)$, owing to the large ratio of the bandwidth to the band centre. The width of the Gaussian is reported as the full width at half maximum:

$$FWHM = \sqrt{8 \ln 2}\ \sigma .$$

The parameters are optimised to minimise the root-mean-square (RMS) error

$$\sqrt{\frac{\int_{IVCT\ band}(A(\nu) - A_{obs}(\nu))^2 \nu^{-2} d\nu}{\int_{IVCT\ band} d\nu}}$$

The observed and fitted spectra for $\mathbf{D_s}$, $\mathbf{D_i}$, $\mathbf{I_s}$, and $\mathbf{I_i}$ are shown in Fig. S4  The fitted parameters and RMS errors (as a percentage of the maximum absorption $A$) are listed in Table S1.

**Table S1.** Parameters in the fit of observed spectra of Prussian blue suspensions $\mathbf{D_s}$, $\mathbf{D_i}$, $\mathbf{I_s}$, and $\mathbf{I_i}$ to a Gaussian function representing IVCT absorption and a power-law baseline representing light scattering.

| sample | time / min | A | $\nu_m$ / eV | $\nu_m$ / cm$^{-1}$ | FWHM / eV | FWHM / cm$^{-1}$ | n | B | % RMS error |
|---|---|---|---|---|---|---|---|---|---|
| $\mathbf{D_s}$ | 0 | 0.570 | 1.734 | 13989 | 0.687 | 5539 | 0 | 0 | 1.65 |
| | 7 | 0.570 | 1.735 | 13997 | 0.684 | 5518 | 0 | 0 | 1.60 |
| | 14 | 0.569 | 1.735 | 13998 | 0.684 | 5521 | 0 | 0 | 1.63 |
| $\mathbf{I_s}$ | 0 | 0.309 | 1.621 | 13075 | 0.928 | 7488 | 2.408 | 0.496 | 1.58 |
| | 7 | 0.316 | 1.604 | 12941 | 0.946 | 7634 | 2.462 | 0.288 | 1.06 |
| | 14 | 0.310 | 1.604 | 12934 | 0.934 | 7530 | 2.533 | 0.210 | 1.10 |
| | 21 | 0.306 | 1.603 | 12931 | 0.924 | 7450 | 2.606 | 0.166 | 1.15 |
| $\mathbf{D_i}$ | 0 | 1.033 | 1.696 | 13677 | 0.800 | 6450 | 0 | 0 | 0.96 |
| | 7 | 1.004 | 1.695 | 13669 | 0.807 | 6513 | 0 | 0 | 0.94 |
| | 14 | 0.974 | 1.690 | 13628 | 0.817 | 6586 | 0 | 0 | 1.09 |
| | 21 | 0.877 | 1.683 | 13575 | 0.826 | 6659 | 0 | 0 | 1.22 |
| $\mathbf{I_i}$ | 0 | 0.805 | 1.544 | 12451 | 1.120 | 9030 | 4.102 | 0.744 | 0.84 |
| | 7 | 0.667 | 1.565 | 12619 | 1.010 | 8145 | 2.420 | 0.446 | 0.66 |
| | 14 | 0.653 | 1.566 | 12629 | 0.994 | 8015 | 2.395 | 0.388 | 0.66 |
| | 21 | 0.642 | 1.569 | 12655 | 0.982 | 7921 | 2.381 | 0.355 | 0.67 |
| | 28 | 0.639 | 1.568 | 12649 | 0.979 | 7899 | 2.424 | 0.327 | 0.66 |
| | 35 | 0.640 | 1.566 | 12633 | 0.982 | 7924 | 2.519 | 0.299 | 0.67 |

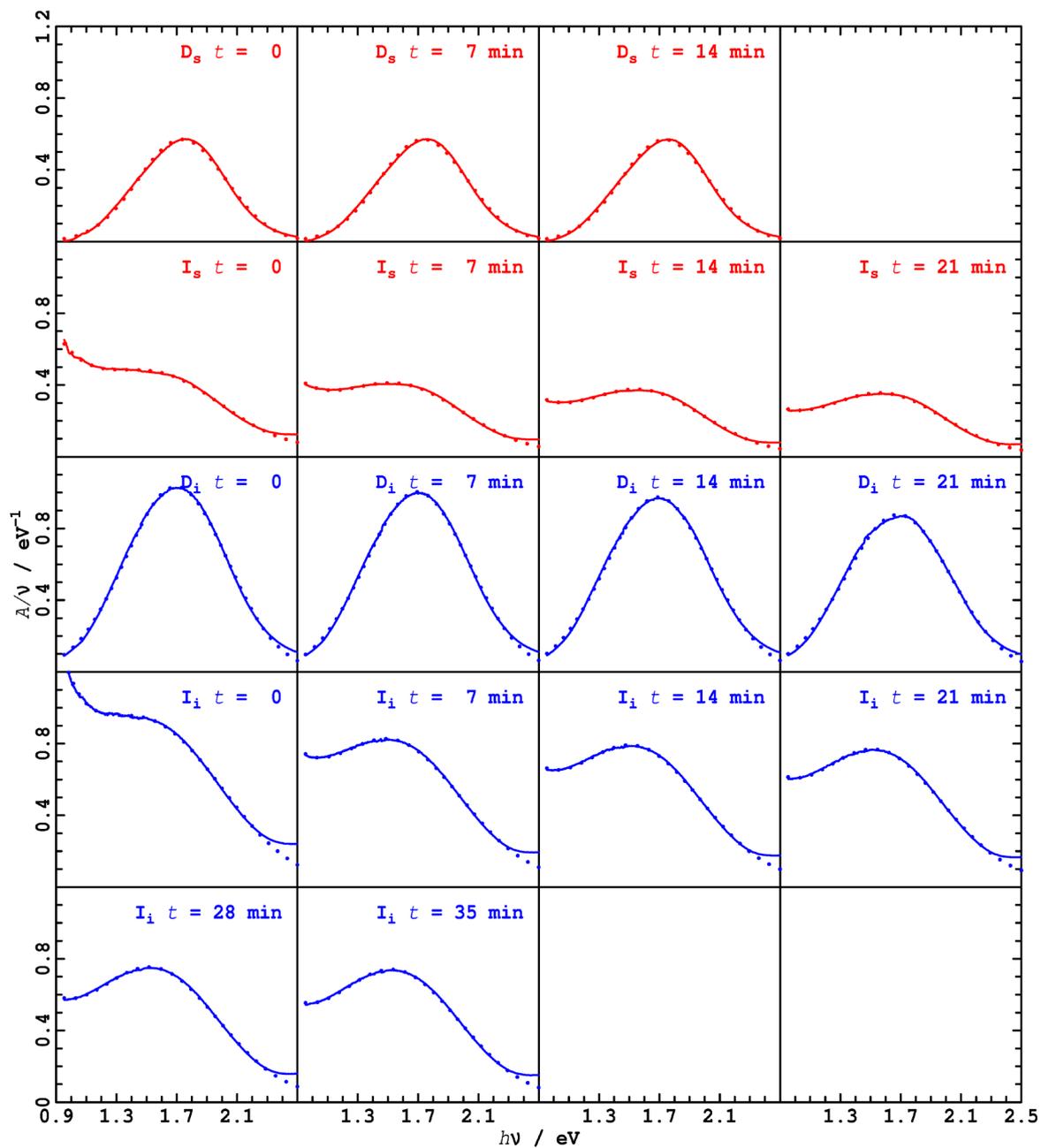

**Fig. S4.** Observed IVCT bandshapes $A/\nu$ for suspensions left standing for varying times $t$ of Prussian blue samples **D$_s$**, **D$_i$**, **I$_s$**, and **I$_i$** (solid lines) are fitted to Gaussian functions with power-law representations of the light scattering (dots). Soluble Prussian blues are in red, insoluble ones in blue.

## b. K$_x$Na$_y$[Fe$_2$(CN)$_{11}$]

The UV-vis-NIR absorption spectrum of the mixed cation salt K$_x$Na$_y$[Fe$_2$(CN)$_{11}$] in water was recorded (Fig. S5).

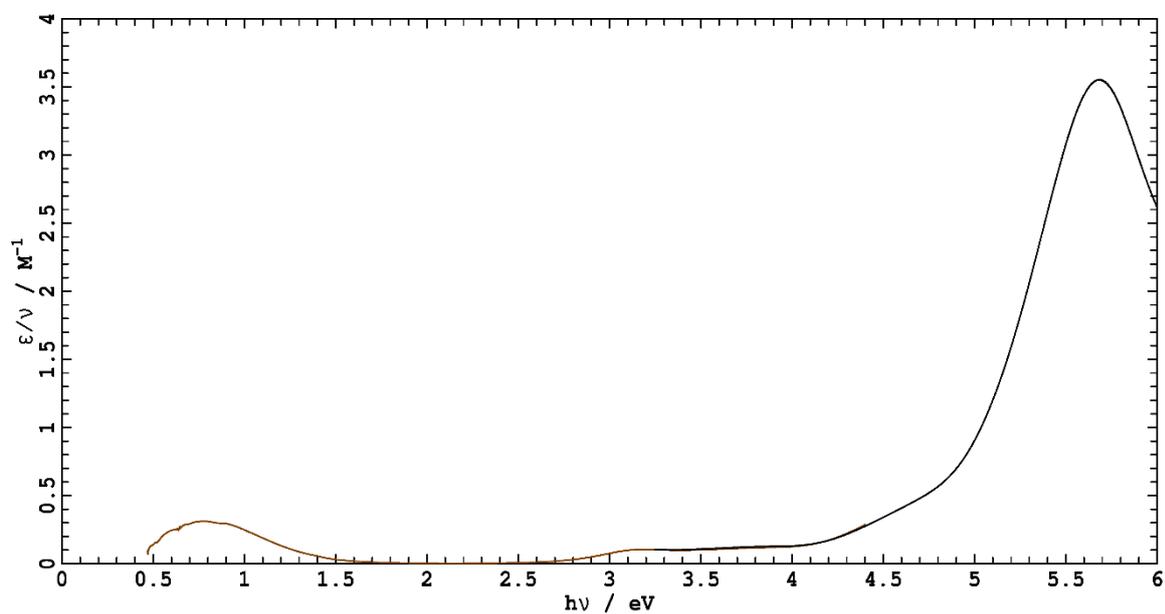

**Fig. S5.** Absorption spectrum of the mixed-valence iron salt K$_x$Na$_y$[{(NC)$_5$Fe$^{II}$}(μ-CN){Fe$^{III}$(CN)$_5$}] in aqueous solution.

# S4. MCD measurements, Magnetic Susceptibility and data analysis.

For **D$_s$** and **D$_i$**, low temperature absorption and MCD spectra were recorded simultaneously on a previously-described(7) laboratory-constructed spectrometer, based on a Spex 0.75 m monochromator. The sample was placed in an Oxford Instruments SM-4 cryostat enabling the temperature to be controlled between 1.8 K and room temperature and magnetic fields of 0 – 5 T to be applied. A 250 W highly stabilised quartz-halogen lamp was used as the light source and a grating blazed at 1000 nm was used for all measurements. Different detectors were used for different wavelength ranges. For the Prussian blue samples, a temperature controlled 5 mm Advanced Photonix silicon avalanche photodiode was used for the 400 – 1100 nm range and an InGaAs photodiode was used for the wavelength range 500 – 1600 nm. For the model mixed-valence dimer, a 1 mm, liquid nitrogen cooled InSb detector system with condensing optics was used in the 700–2500 nm range. Measurements were made in sections with order sorting filters where necessary.

A concentrated aqueous suspension of Prussian blue was mixed 50:50 with glycerol. The mixture was then pressed between two 12 mm quartz windows with a 50 μm mylar spacer. The K$_y$Na$_x$[Fe$_2$(CN)$_{11}$] model complex was dissolved in a small quantity of water and then diluted 20:80 with glycerol before being pressed between 12 mm windows in the same fashion as that of the Prussian blue samples. For **P$_i$**, MCD spectra were measured at 10 K in a PVA film at 5 T. The MCD signal amplitudes for K$_y$Na$_x$[Fe$_2$(CN)$_{11}$] and the Prussian blue **D$_s$** measured as a function of field and temperature are shown in Fig. S6 along with model simulations.

The magnetic properties of the ground state of a chromophore are imbedded in the field and temperature dependence of its MCD spectrum(8). In the simplest case of an isolated magnetic doublet, MCD amplitude measurements, when scaled to $B/2kT$, fall on a Brillouin curve. This accounts for the thermal population differences of the two Zeeman sub-levels of the doublet, with each sub-level having transitions with equal and opposite circular polarisation. This simple case indeed applies to the [Fe$_2$CN$_{11}$]$^{6-}$ dimer data in Fig. S6 where $S$ = 1/2 and $g$ = 1.8 accounts *quantitatively* for the data, the parameters being fully consistent with EPR data.

A heuristic analysis can be extended to the case for $S$ > 1/2 by using the spin Hamiltonian model of Neese and Solomon(9). This model assumes that the material can be described as a "molecular-crystal" in which spins *within* the "molecule" are strongly coupled, with no spin-spin couplings *between* "molecules". Beyond this, it is agnostic with respect to the details of the chromophore involved and reports *parametric g* values and zero field splittings (ZFSs). Rather than falling on a universal curve, the field dependence at each temperature shows a nesting behaviour with each higher temperature curve falling away from the lowest temperature curve. The extent of this effect can be used to estimate the ZFSs. As is seen in Fig. S6, the MCD nesting behaviour of Prussian blue can be simulated using a value of $S$ = ~10/2 and a relatively small ZFS of the order of 0.5 cm$^{-1}$. Nevertheless, we note that the nesting behaviour seen is not quantitatively accounted for by this model for Prussian blue. The spin Hamiltonian model has been extended to the case of multiple coupled spins(10), but there appears to be no good structural reason to apply such a model to Prussian blue as it is not clear how to divide the magnetic interactions into the "weak" and "strong" classes required for application of this molecular-crystal approach. In addition, the temperature dependence of the purely magnetic component of the neutron diffraction data below the Curie temperature(11) exclude the possibility that, e.g., strongly ferromagnetically coupled Fe(III) dimers, as a result of $S = 10/2$ implies, can be identified within Prussian blue.

Bulk magnetisation measurements are clear in indicating that the spin state of the Prussian blue materials above 5 K is $S = 5/2$ with small ZFS(12). We have repeated room temperature magnetic susceptibility measurements on our samples using a Sherwood MSB Mk1 magnetic susceptibility balance (Table S2) and have confirmed previous results in being consistent with the Prussian blue being made up of locally high spin Fe(III) ($^6S$, $S = 5/2$) and locally low spin Fe(II) ($^1A$, $S = 0$) in all samples.

The close similarity of MCD and MCD saturation data for all Prussian blue samples point to there being characteristic chromophore(s) responsible for the IVCT absorption that are not significantly affected by the disorder and variability of this archetypal coordination polymer. Modelling of the MCD behaviour would involve a detailed calculation of both electric and magnetic dipole moments associated with the chromophores involved in IVCT excitations in Prussian blue. We do note that in spectra of a series of mixed valence dimers(13) in which the bridging ligands between the Ru(II) and Ru(III) ions is changed from chloride to bromide, leads to the sign of the MCD of the IVCT band *reversing*. This sensitivity of MCD to small changes may well relate to the structure in the MCD of Prussian blue as being assigned to energetically distinct ligand isomer IVCT bands.

**Table S2.** Room temperature magnetic susceptibilities of all Prussian blue samples studied in this work.

| Sample | Formula used | Molecular weight (g/mol) | Mass susceptibility (cm$^3$/g) | Molar susceptibility (cm$^3$/mol) | Total $\mu_{eff}$ | $\mu_{eff}$ per Fe(III) |
|---|---|---|---|---|---|---|
| $P_i$ | Fe$_4$[Fe(CN)$_6$]$_3$.6H$_2$O | 967 | 5.51 x 10$^{-5}$ | 0.053 | 11.2 | 5.6 |
| $D_s$ | Fe$_4$[Fe(CN)$_6$]$_3$.6H$_2$O | 967 | 4.66 x 10$^{-5}$ | 0.045 | 10.3 | 5.2 |
| $D_i$ | Fe$_4$[Fe(CN)$_6$]$_3$.6H$_2$O | 967 | 4.56 x 10$^{-5}$ | 0.043 | 10.1 | 5.1 |
| $I_s$ | Fe$_4$[Fe(CN)$_6$]$_3$.6H$_2$O | 967 | 5.14 x 10$^{-5}$ | 0.050 | 10.8 | 5.4 |
| $I_i$ | Fe$_4$[Fe(CN)$_6$]$_3$.6H$_2$O | 967 | 5.11 x 10$^{-5}$ | 0.049 | 10.8 | 5.4 |

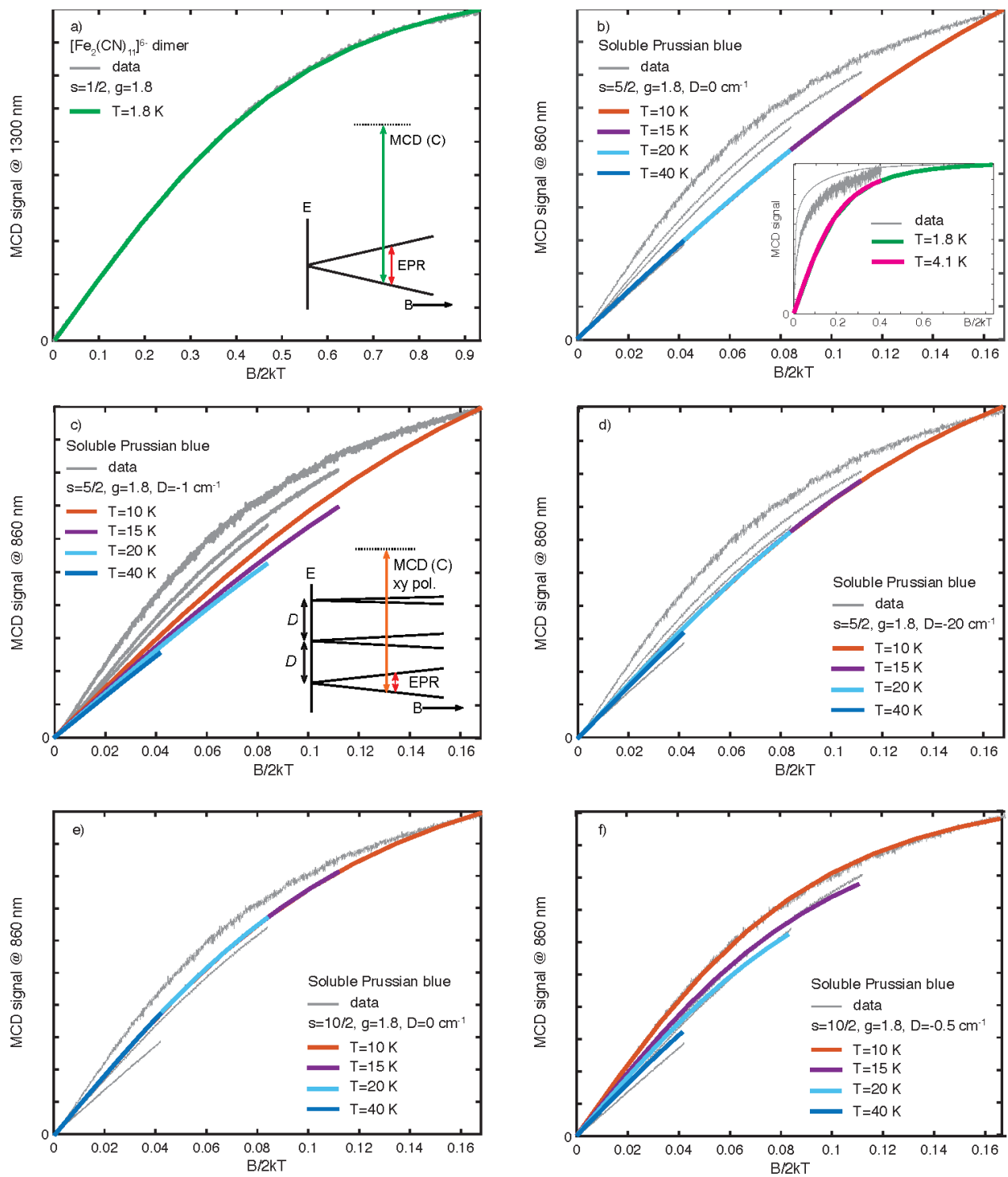

**Fig. S6.** MCD saturation modelling and data for soluble Prussian blue $D_s$ and the model dimeric ion $[Fe_2(CN)_{11}]^{6-}$. The model was based on that of Neese and Solomon(9) and implemented using Matlab.

# S5. CW EPR measurements.

## a. Methods

Samples of soluble Prussian blue $I_s$ and insoluble Prussian blue $I_i$ were measured as powder samples by continuous Wave (CW) EPR measurements performed at 10 K on a Bruker E500 X-band spectrometer (~9.8 GHz) equipped with an ER4122 SHQ resonator. Spectra were recorded with a field modulation amplitude of 20 G and a power of 0.047 mW. The Prussian blue model ion $[Fe_2(CN)_{11}]^{6-}$ was also measured as a frozen solution spectrum at 10 K, in ~2 mM concentration sonicated in a 50/50 MeOH/H$_2$O solvent mix. A solution of K$_3$[Fe(CN)$_6$] (potassium ferricyanide), prepared in the same way, was also measured to confirm that the dimer does not decompose when suspended MeOH/H$_2$O. Simulations were performed using the solid-state CW EPR program *pepper* in the EasySpin package(14) in MATLAB, and were optimized using a Nelder-Mead minimization algorithm (see Fig. 4).

## b. Results

The observed X-band EPR spectra are shown in Fig. S7. The powder X-band spectra of the Prussian Blue samples $I_s$ and $I_i$ are characterised by broad unstructured signals. In contrast, the spectrum of the $[Fe_2(CN)_{11}]^{6-}$ model ion containing Fe(II) and Fe(III) measured as a powdered sample is structured, and can be assigned to low spin Fe(III), with the feature at ~170 mT most likely corresponding to a small high spin component. The Fe(III) assignment is confirmed by the analogous solution spectrum and its corresponding simulation shown in Fig. 4b; i.e. the spectrum can be modelled by an $S = 1/2$ spin system dominated by *g*-anisotropy. To account for the linewidth, both a Gausssian linewidth (1 mT) and an H-strain was included, the latter of which accounts for co-contributions to line broadening by unresolved hyperfine splittings. The specific simulation parameters used to model the dimer solution spectrum are detailed in Table S3.

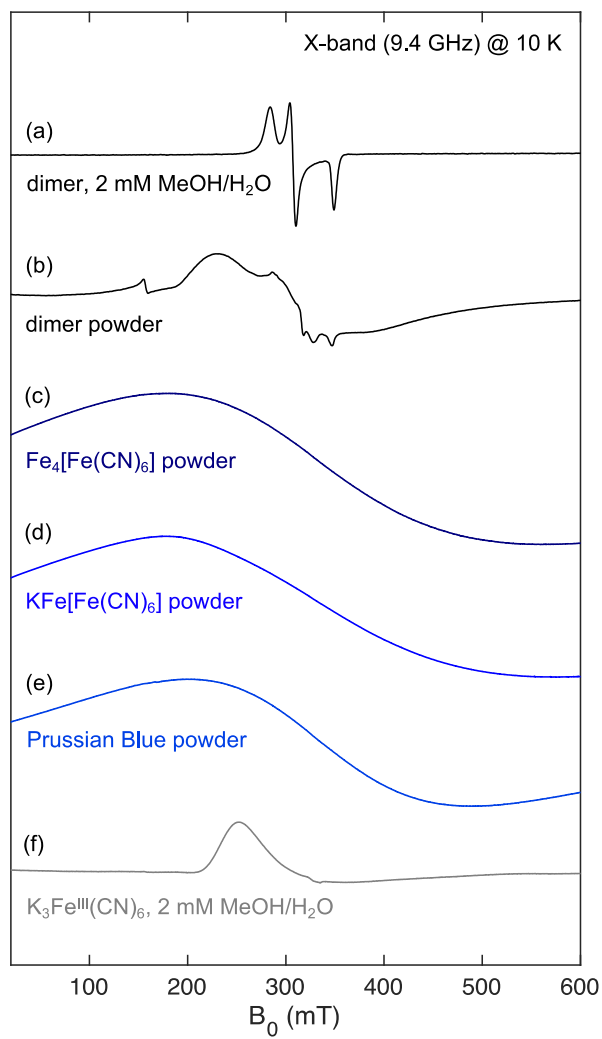

**Fig. S7.** X-band CW frozen powder spectra of Prussian blue samples $\mathbf{I_s}$ and $\mathbf{I_i}$ and the model mixed-valence ion $[Fe_2(CN)_{11}]^{6-}$, as well as analogous solution spectra for this ion and for potasssium ferricyanide $K_3[Fe^{III}(CN)_6]$, measured at 10 K.

**Table S3.** Frozen solution spectrum simulation parameters for $[Fe_2(CN)_{11}]^{6-}$.

| Parameter | Modelled Value |
|---|---|
| g-tensor [$g_{xx}$ $g_{yy}$ $g_{zz}$] | [2.36  2.18.  1.92] |
| Linewidth (mT) | 1.8 |
| g-strain [x  y  z] | [0.0243  0.01542.  0.04999] |

# S6. Pulse EPR Measurements @ X-band

## a. Methods

HYSCORE (Hyperfine Sublevel CORrElation spectroscopy) spectra were recorded at X-band at 10 K on a Bruker E580 X-band spectrometer (~9.8 GHz) equipped with an ER4118 XMD-5-W1 pulse resonator. The same frozen solution of the $K_xNa_y[Fe_2(CN)_{11}]$ model complex was used for both the CW measurements previously reported and these pulsed measurements.

## b. Results

The structure of the $[Fe_2(CN)_6]^{6-}$ model ion involves a μ-CN bridge between its Fe(III) and Fe(II) centres. The 3-pulse ESEEM and HYSCORE measurements were used to measure the nitrogen couplings from the CN ligands to identify the orientation (κ-CN vs κ-NC) of the μ-CN bridge towards each iron centre.

The HYSCORE experiment is a 2D version of the 3-pulse ESEEM experiment and discriminates between interactions in the strong and weak coupling regimes. Weakly coupled nuclei appear at the nuclear Larmor frequency in the (+, +) quadrant (upper right), split by the hyperfine frequency about the antidiagonal. Strongly coupled nuclei appear in the (+, -) quadrant (upper left) at half the hyperfine frequency ($A/2$), split by twice the Larmor frequency $2\upsilon_N$ about the diagonal.

Figure S8 shows the HYSCORE spectrum measured at each $g$ value, i.e., turning point of the field-sweep electron-spin-echo (ESE) spectrum (insert, Fig. S8). The spectra are simulated in Fig. S9 using the *EasySpin* packagae *saffron*. Based on parameters listed in Table S4. The simulations predict and interpret all key spectral features.

In each spectrum, there is a peak and broad correlation ridge at ~15 MHz in the (+,+) quadrant corresponding to a dipolar coupled proton, most likely arising from a water ligand. Similarly an intense correlation is observed at the $^{13}C$ Larmor frequency position, which is assigned to the carbon-bound CN ligands. Signals that can be assigned to a $^{14}N$ are present in both the (+,+) and (+,-) quadrants, suggesting an intermediate coupling regime in which the nitrogen hyperfine coupling is of the order of the $^{14}N$ Larmor frequency. In the (+,+) quadrant indicative of weak coupling, there is some signal intensity at ~1.1 MHz which is the single quantum $^{14}N$ transition frequency, but a larger correlation ridge appears at ~2.2 MHz, assigned to the double quantum transition. These splittings are on the order of ~1.4 and ~2.8 MHz respectively.

In the (+,-) quadrant, there is a two correlation ridges, centered at 0.8 and 3.6 MHz, with splittings broadly corresponding to two times $\upsilon_N(^{14}N)$. The (+,-) correlation patterns in particular are strongly diagnostic of Fe(III)-bound and therefore strongly coupled nitrogens, indicating a bridged structure where the nitrogen is bound via the Fe(III) and the carbon is bound to the Fe(II). The calculated isotropic coupling constants listed in Table S5 and shown pictorially in Fig. S10 confirm this analysis. These calculations were performed using Gaussian16 and are done at both the CAM-B3LYP and SAC-CI level. Both methods predict that the ground state is $[\{NC\}_5Fe^{II}\}(\mu\text{-}CN)\{Fe^{III}(CN)_5\}]$, but allow access to the alternative possibility through modelling of its intervalence charge-transfer transition and the associated excited state $([\{NC\}_5Fe^{III}\}(\mu\text{-}CN)\{Fe^{II}(CN)_5\}]^*)$.

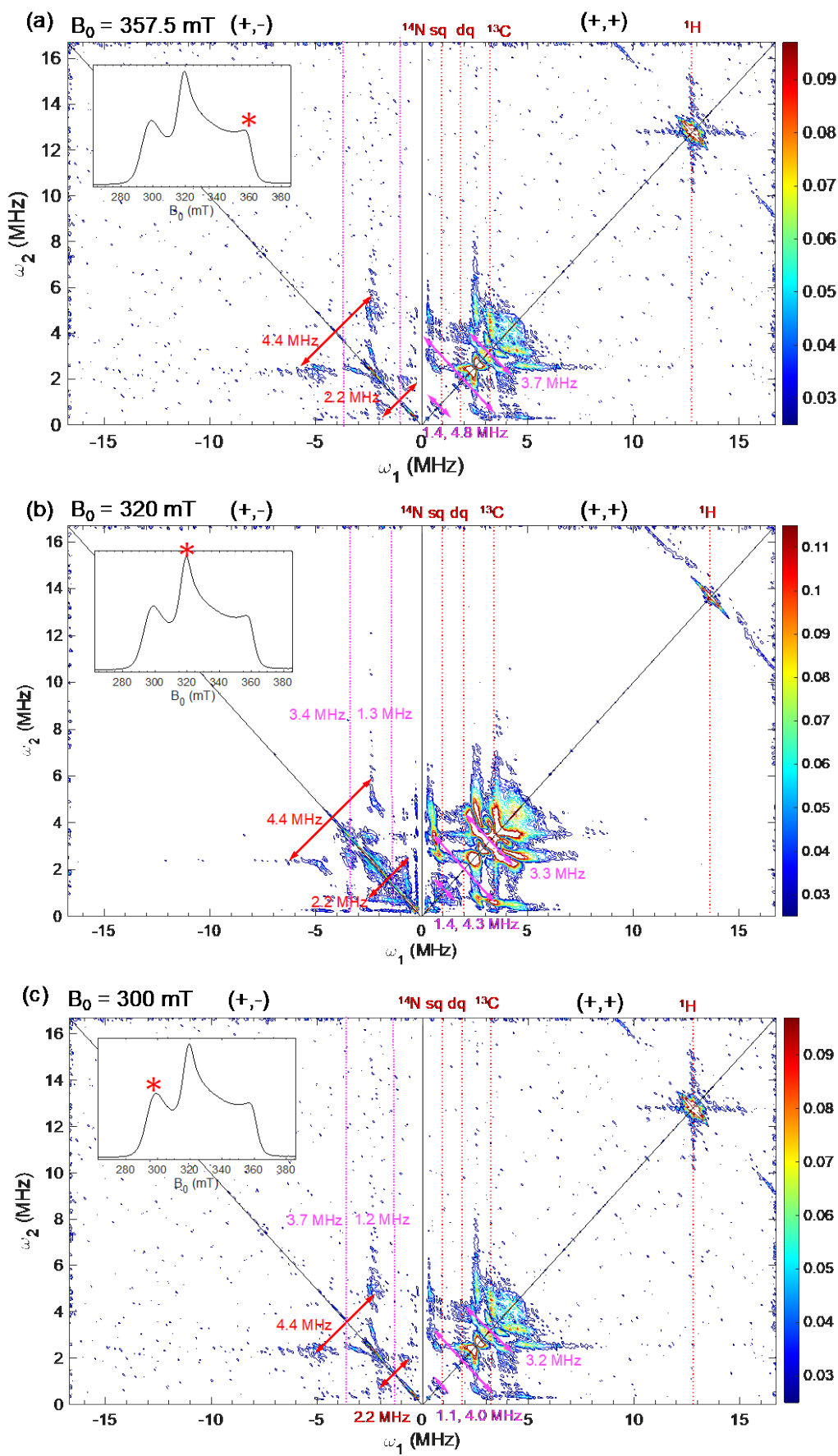

**Fig. S8.** HYSCORE spectra of $[Fe_2(CN)_{11}]^{6-}$.

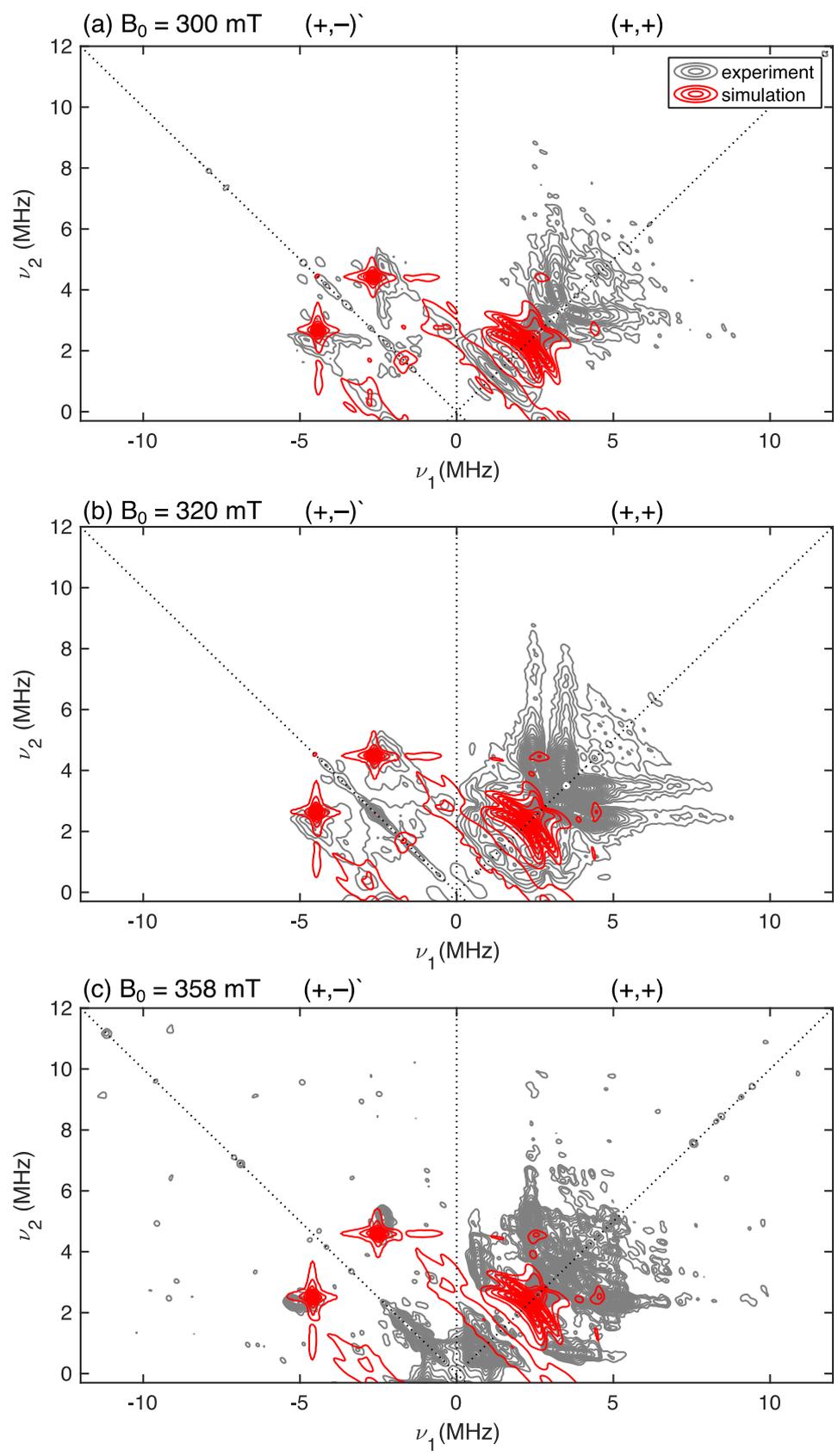

**Fig. S9**. Experimental HYSCORE data (grey) vs simulations (red) performed in. *EasySpin* using the parameters detailed in Table S4.

**Table S4.** HYSCORE simulation parameters of the nitrogen region of the experimental data. Simulations were performed using the *EasySpin* packagae *saffron*. The initial values for simulation were based on the work of Vinck and Van Doorslaer(15).

| g-tensor | $^{14}$N dipolar coupling (MHz) | $^{14}$N isotropic coupling (MHz) | $eq^2Q/h$ (MHz) | $\eta$ |
|---|---|---|---|---|
| [2.36  2.18  1.92] | [4.8  4.8  -9.6] | 2.0 | 2.8 | 0.14 |

**Table S5.** Calculated isotropic coupling for $[Fe^1(CN)_5 - CN - Fe^2(CN)_5]^{6-}$ in both its $Fe^{2+}(CN^-)_5 - CN^- - Fe^{3+}(CN^-)_5$ and $Fe^{3+}(CN^-)_5 - CN^- - Fe^{2+}(CN^-)_5$ valence states.

| Atom | | | $Fe^{2+}(CN^-)_5 - CN^- - Fe^{3+}(CN^-)_5$ | | $Fe^{3+}(CN^-)_5 - CN^- - Fe^{2+}(CN^-)_5$ | |
|---|---|---|---|---|---|---|
| | | | CAM-B3LYP | SAC-CI | CAM-B3LYP | SAC-CI |
| $Fe^1(CN)_5$ | N | | -0.6 | 0.2 | 0.0 | 0.0 |
| | | | 0.9 | -0.1 | 0.0 | 0.0 |
| | | | 0.9 | 0.1 | 0.0 | 0.0 |
| | | | -0.6 | 0.1 | 0.0 | 0.0 |
| | | | 1.2 | -0.1 | 0.0 | 0.0 |
| | C | | **-36.0** | **-10.0** | -0.1 | -0.2 |
| | | | **-33.4** | **-11.4** | 1.3 | -0.2 |
| | | | **-33.4** | **-10.7** | 1.3 | 0.3 |
| | | | **-36.0** | **-10.7** | -0.1 | 0.3 |
| | | | **-41.6** | **-10.0** | -0.2 | -0.2 |
| bridge | N | | **-6.4** | **-0.5** | 1.1 | 0.4 |
| | C | | 3.1 | -1.0 | **-39.1** | **-13.3** |
| $Fe^2(CN)_5$ | N | | 0.0 | 0.0 | 0.8 | 0.0 |
| | | | 0.0 | 0.0 | -0.6 | 0.0 |
| | | | 0.0 | 0.0 | 0.8 | 0.0 |
| | | | 0.0 | 0.0 | 0.8 | 0.2 |
| | | | 0.0 | 0.0 | -0.6 | 0.0 |
| | C | | -0.1 | -1.0 | **-36.3** | **-13.4** |
| | | | 1.3 | -1.0 | **-36.4** | **-13.4** |
| | | | -0.1 | -1.2 | **-38.3** | **-12.9** |
| | | | 1.3 | -1.0 | **-36.4** | **-13.4** |
| | | | -0.1 | -1.0 | **-38.3** | **-13.4** |

In addition, the observed HYSCORE spectra agree well with that reported for Fe(III)-ligated nitrogen atoms(16) in the bound states of the biotin enzyme and its intermediates to an iron-sulfur bridge during enzymatic reactions catalysing C-S bond formation. The nitrogen correlation peaks appear in these systems at almost exactly the same positions in both quadrants, with splittings between 2-6 MHz. Similar hyperfine splittings and HYSCORE correlation patterns are shown in the work of Vinck and Van Doorslaer on EPR studies of low spin Fe(III) porphyrin complexes(15).

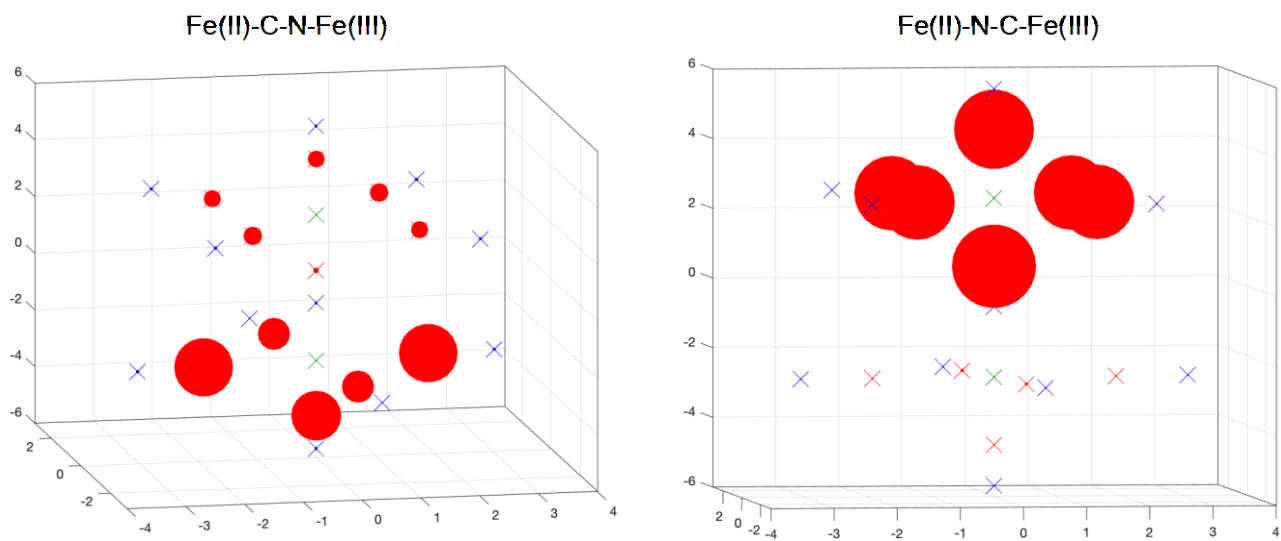

**Fig. S10.** Calculated % Isotropic Fermi Contact coupling on each atom of $[Fe_2(CN)_{11}]^{6-}$.

## S7. Computational modelling.

### a. Methods

All calculations on model ions were performed using the Gaussian16 suite of programs.(17) The SAC-CI,(18) DFT, and TDDFT calculations were performed using the LANL2DZ(19) basis set for Fe and otherwise 6-31+G*(20) for geometry optimisations and 6-31G(20) for TDDFT calculations. Test TDDFT calculations using the MDF10 basis set(21) for Fe and/or 6-31G*(20) or 6-311+G*(20) revealed that basis set size is not a major factor determining calculation quality. The "level one" approximation was used for SAC-CI. The PCM model(22) for water was used to generate the SCRF for $[Fe_2(CN)_{11}]^{6-}$ in solution. Special methods are used to describe the electrostatic potential for model compounds intended to mimic properties of Prussian blue, see later.

      All calculations on periodic solids were performed using VASP(23). PAW pseudopotentials(24) were used, HIGH precision, and thermal broadening of 0.05 eV for ground-state calculations and geometry optimisations. Excited-state energies were obtained by freezing the electron occupancy to give the HOMO-LUMO excitation. Geometry optimisations, band gaps, and excitation energies were performed on cubic unit cells of length 10.16 Å using 2 × 2 × 2 k-points; analogously expanded unit cells were also considered at their Γ-point and verified to give the same results for band gaps and excitation energies. CAM-B3LYP was implemented into VASP as described elsewhere(25). Additional calculations using the HSE06 functional(26) were also performed. Optimised Cartesian coordinates for the model compounds and solids are listed in Supporting Information Section S9. CAM-B3LYP transition moments for Prussian blue were evaluated by the VASPKit code(27), and the BSE calculations were performed(28) using the CAM-B3LYP+BSE and HSE06+BSE.

### b. Results for model compounds

The structure and relative energy of κ-CN and κ-NC ligation isomers of $[Fe_2(CN)_{11}]^{6-}$, in both low-spin and high-spin states and staggered and eclipsed conformations, were modelled using the CAM-B3LYP density functional using a self-consistent reaction-field (SCRF) solvent model for aqueous solution (Table S6). Unsurprisingly, the linkage isomer with all peripheral ligands bound to $Fe^{2+}$ via C (i.e. κ-CN) was predicted to be the most stable by a large amount, 38 kcal mol$^{-1}$ (1.6 eV), with the κ-NC forms being the highest energy structures. Whereas the energy differences between staggered and eclipsed conformations were predicted to be small, the ground state was predicted to be 24 kcal mol$^{-1}$ (1.0 eV) more stable as low-spin than as high-spin. These predictions clearly align with the key observed features of linkage isomerism and overall spin, but prediction of the location of the spin and the properties of the IVCT band is hampered by the anticipated $^2E$ nature of the ground state. Such states do not obey the Kohn-Sham theorem and are sensitive to static electron correlations not accounted for in standard DFT. Indeed, time-dependent DFT (TD-DFT) calculations for the $^2A_2$ state indicate that $^2E$ lies 0.26 eV lower in energy, in good agreement with the experimental result obtained using Griffith's theory of 0.28 eV. Setting the electron occupancy to one of the $^2B_1$ or $^2B_2$ components of $^2E$ and forcing the unpaired spin to localise on alternative Fe ions gives an energy difference of just 0.03 eV between the linkage isomers [{(NC)$_5$Fe$^{II}$}(μ-CN){Fe$^{III}$(CN)$_5$}] and [{(NC)$_5$Fe$^{II}$}(μ-NC){Fe$^{III}$(CN)$_5$}]. That these two states should have so similar energies is qualitatively consistent with the observed very low origin energy of the IVCT band of $h\nu_{00} = 0.33$ eV

Table S6. Relative energies calculated using CAM-B3LYP with the MDF10 basis set for Fe and 6-311+G* otherwise and the PCM solvation model for ligation isomers of $[Fe_2(CN)_{11}]^{6-}$ of form $FeX_5 - CN - FeY_5$.

| X | Y | Conformer | Fe$^{III}$ Spin | ΔE / kcal mol$^{-1}$ |
|---|---|---|---|---|
| κ-CN | κ-CN | staggered C$_2$ | low | [0] |
| κ-CN | κ-CN | staggered C$_{4v}$ | low | 0.7 |
| κ-CN | κ-CN | eclipsed | low | 0.3 |
| κ-CN | κ-CN | staggered | high | 24 |
| κ-CN | κ-CN | eclipsed | high | 25 |
| κ-CN | κ-NC | staggered | high | 38 |
| κ-CN | κ-NC | eclipsed | high | 38 |
| κ-CN | κ-NC | eclipsed | low | 47 |
| κ-CN | κ-NC | staggered | low | 55 |
| κ-NC | κ-NC | staggered | high | 78 |
| κ-NC | κ-NC | eclipsed | high | 79 |
| κ-NC | κ-NC | eclipsed | low | 99 |
| κ-NC | κ-NC | staggered | low | 100 |

As an alternative to DFT, the SAC-CI coupled-cluster *ab initio* approach was also applied to calculate the structure and spectroscopy of the mixed-valence (Fe$^{II}$-Fe$^{III}$) ion $[Fe_2(CN)_{11}]^{6-}$. In this approach, the electronic structure and solvent associated reaction field are calculated for the analogous homovalent Fe$^{II}$-Fe$^{II}$ ion $[Fe_2(CN)_{11}]^{7-}$, and then the ground state and excited states of $[Fe_2(CN)_{11}]^{6-}$ obtained by ionisation. When the intrinsic Fe$^{II}$-Fe$^{II}$ reaction field is replaced with the CAM-B3LYP reaction field for the Fe$^{II}$-Fe$^{III}$ mixed-valence ion $[(NC)_5Fe^{II}$}(μ-CN){Fe$^{III}$(CN)$_5$}]$^{6-}$, an IVCT band is predicted at $h\nu_v = 0.34$ eV with a transition moment of $M = 1.9$ eÅ. This energy is less than that observed by 0.52 eV, and the transition moment is double that observed. Such errors arise as the calculated properties are very sensitive to the treatment of the solvent reaction field. Using the native reaction field coming from the $Fe^{II}$-$Fe^{II}$ state, the transition energy becomes -0.96 eV (*i.e.*, [(NC)$_5$Fe$^{II}$}(µ-NC){Fe$^{III}$(CN)$_5$}]$^{6-}$, is incorrectly predicted to be the ground state of the ion), falling further to -2.21 eV if the reaction field for this inverted configuration is used; the associated transition moments fall to 0.40 eÅ and 0.16 eÅ, respectively. Given these uncertainties, the coupled-cluster based SAC-CI approach is seen to produce a realistic description of the spectroscopy of the model ion in solution.

### c. External potential for modelling model compounds mimicking Prussian blue

The electric potential difference between neighbouring Fe(II) and Fe(III) atoms in a cell of Prussian blue arising from its external environment was estimated by performing a Bader charge analysis(29) of the electronic structure from the CAM-B3LYP calculations on $Fe_3^{2+}Fe_4^{3+}(CN^-)_{18}(H_2O)_6$ and $Fe^{2+}Fe^{3+}(CN^-)_6 \cdot K^+$. The Bader charges are listed in Tables S6-S7. The deduced average potential difference between Fe(II) and Fe(III) from an infinite sum of images of these charges is 0.76 V for $Fe_3^{2+}Fe_4^{3+}(CN^-)_{18}(H_2O)_6$ and 0.60 V for $Fe^{2+}Fe^{3+}(CN^-)_6 \cdot K^+$. These differences were then fitted to a model for the model compounds in which a point charge is placed at the location of every nearest-neighbour Fe atom. To match the potential difference determined from the solid, charges of 0.2 e adjacent to Fe(III) and 0.8 e adjacent to Fe(II) were determined. The resulting charge sets used for each model compound are listed along with the coordinates in Section S9.

**Table S7.** Bader charges for each atom in $Fe_3^{2+}Fe_4^{3+}(CN^-)_{18}(H_2O)_6$.

| Atomic Number | x | y | z | charge / e |
|---|---|---|---|---|
| 26 | 5.08 | 5.08 | 5.08 | 1.890 |
| 26 | 5.08 | 0 | 0 | 1.840 |
| 26 | 0 | 0 | 5.08 | 1.840 |
| 26 | 0 | 5.08 | 0 | 1.840 |
| 26 | 0 | 5.08 | 5.08 | 0.727 |
| 26 | 5.08 | 0 | 5.08 | 0.727 |
| 26 | 5.08 | 5.08 | 0 | 0.727 |
| 8 | 7.206844 | 0 | 0 | -1.541 |
| 8 | 2.953156 | 0 | 0 | -1.541 |
| 8 | 0 | 7.206844 | 0 | -1.541 |
| 8 | 0 | 2.953156 | 0 | -1.541 |
| 8 | 0 | 0 | 7.206844 | -1.541 |
| 8 | 0 | 0 | 2.953156 | -1.541 |
| 7 | 2.012462 | 0 | 5.08 | -1.531 |
| 7 | 0 | 2.014667 | 5.08 | -1.515 |
| 7 | 2.014667 | 5.08 | 0 | -1.515 |
| 7 | 0 | 5.08 | 2.012462 | -1.531 |
| 7 | 3.045673 | 5.08 | 5.08 | -1.428 |
| 7 | 0 | 8.145333 | 5.08 | -1.515 |
| 7 | 0 | 5.08 | 8.147538 | -1.531 |
| 7 | 5.08 | 2.012462 | 0 | -1.531 |
| 7 | 5.08 | 0 | 2.014667 | -1.515 |
| 7 | 5.08 | 3.045673 | 5.08 | -1.428 |
| 7 | 8.147538 | 0 | 5.08 | -1.531 |
| 7 | 5.08 | 0 | 8.145333 | -1.515 |
| 7 | 5.08 | 5.08 | 3.045673 | -1.428 |
| 7 | 8.145333 | 5.08 | 0 | -1.515 |
| 7 | 5.08 | 8.147538 | 0 | -1.531 |
| 7 | 7.114327 | 5.08 | 5.08 | -1.428 |
| 7 | 5.08 | 7.114327 | 5.08 | -1.428 |
| 7 | 5.08 | 5.08 | 7.114327 | -1.427 |
| 6 | 3.178332 | 0 | 5.08 | 0.962 |
| 6 | 0 | 3.17756 | 5.08 | 0.955 |
| 6 | 3.17756 | 5.08 | 0 | 0.955 |
| 6 | 0 | 5.08 | 3.178332 | 0.962 |
| 6 | 1.893926 | 5.08 | 5.08 | 0.888 |
| 6 | 0 | 6.98244 | 5.08 | 0.955 |
| 6 | 0 | 5.08 | 6.981668 | 0.962 |
| 6 | 5.08 | 3.178332 | 0 | 0.962 |
| 6 | 5.08 | 0 | 3.17756 | 0.955 |
| 6 | 5.08 | 1.893926 | 5.08 | 0.888 |
| 6 | 6.981668 | 0 | 5.08 | 0.962 |
| 6 | 5.08 | 0 | 6.98244 | 0.955 |
| 6 | 5.08 | 5.08 | 1.893926 | 0.888 |
| 6 | 6.98244 | 5.08 | 0 | 0.955 |
| 6 | 5.08 | 6.981668 | 0 | 0.962 |
| 6 | 8.266074 | 5.08 | 5.08 | 0.888 |
| 6 | 5.08 | 8.266074 | 5.08 | 0.888 |
| 6 | 5.08 | 5.08 | 8.266074 | 0.888 |
| 1 | 7.712283 | 9.342252 | 0 | 0.805 |
| 1 | 7.712283 | 0.817748 | 0 | 0.805 |
| 1 | 2.447717 | 9.342252 | 0 | 0.805 |
| 1 | 2.447717 | 0.817748 | 0 | 0.805 |
| 1 | 0 | 7.712283 | 9.342252 | 0.805 |
| 1 | 0 | 7.712283 | 0.817748 | 0.805 |
| 1 | 0 | 2.447717 | 9.342252 | 0.805 |
| 1 | 0 | 2.447717 | 0.817748 | 0.805 |
| 1 | 9.342252 | 0 | 7.712283 | 0.805 |
| 1 | 0.817748 | 0 | 7.712283 | 0.805 |
| 1 | 9.342252 | 0 | 2.447717 | 0.805 |
| 1 | 0.817748 | 0 | 2.447717 | 0.805 |
| | | | | 0.000 |

**Table S8.** Bader charges for each atom in $Fe^{2+}Fe^{3+}(CN^-)_6 \cdot K^+$.

| Atomic Number | x | y | z | charge / e |
|---|---|---|---|---|
| 26 | 5.08 | 0 | 0 | 1.894 |
| 26 | 0 | 5.08 | 0 | 1.894 |
| 26 | 0 | 0 | 5.08 | 1.894 |
| 26 | 5.08 | 5.08 | 5.08 | 1.894 |
| 26 | 0 | 0 | 0 | 0.627 |
| 26 | 5.08 | 5.08 | 0 | 0.627 |
| 26 | 5.08 | 0 | 5.08 | 0.627 |
| 26 | 0 | 5.08 | 5.08 | 0.627 |
| 19 | 2.54 | 2.54 | 2.54 | 0.932 |
| 19 | 2.54 | 7.62 | 7.62 | 0.932 |
| 19 | 7.62 | 2.54 | 7.62 | 0.932 |
| 19 | 7.62 | 7.62 | 2.54 | 0.932 |
| 7 | 3.035605 | 0 | 0 | -1.425 |
| 7 | 0 | 3.035605 | 0 | -1.425 |
| 7 | 5.08 | 2.044395 | 0 | -1.425 |
| 7 | 2.044395 | 5.08 | 0 | -1.425 |
| 7 | 0 | 0 | 3.035605 | -1.425 |
| 7 | 5.08 | 5.08 | 3.035605 | -1.425 |
| 7 | 5.08 | 0 | 2.044395 | -1.425 |
| 7 | 0 | 5.08 | 2.044395 | -1.425 |
| 7 | 2.044395 | 0 | 5.08 | -1.425 |
| 7 | 0 | 2.044395 | 5.08 | -1.425 |
| 7 | 5.08 | 3.035605 | 5.08 | -1.425 |
| 7 | 3.035605 | 5.08 | 5.08 | -1.425 |
| 7 | 7.124395 | 0 | 0 | -1.425 |
| 7 | 7.124395 | 5.08 | 5.08 | -1.425 |
| 7 | 8.115605 | 5.08 | 0 | -1.425 |
| 7 | 8.115605 | 0 | 5.08 | -1.425 |
| 7 | 0 | 7.124395 | 0 | -1.425 |
| 7 | 5.08 | 7.124395 | 5.08 | -1.425 |
| 7 | 5.08 | 8.115605 | 0 | -1.425 |
| 7 | 0 | 8.115605 | 5.08 | -1.425 |
| 7 | 0 | 0 | 7.124395 | -1.425 |
| 7 | 5.08 | 5.08 | 7.124395 | -1.425 |
| 7 | 5.08 | 0 | 8.115605 | -1.425 |
| 7 | 0 | 5.08 | 8.115605 | -1.425 |
| 6 | 1.878604 | 0 | 0 | 0.849 |
| 6 | 0 | 1.878604 | 0 | 0.849 |
| 6 | 5.08 | 3.201396 | 0 | 0.849 |
| 6 | 3.201396 | 5.08 | 0 | 0.849 |
| 6 | 0 | 0 | 1.878604 | 0.850 |
| 6 | 5.08 | 5.08 | 1.878604 | 0.850 |
| 6 | 5.08 | 0 | 3.201396 | 0.849 |
| 6 | 0 | 5.08 | 3.201396 | 0.849 |
| 6 | 3.201396 | 0 | 5.08 | 0.849 |
| 6 | 0 | 3.201396 | 5.08 | 0.849 |

| | | | | | |
|---|---:|---:|---:|---:|---:|
| 6 | | 5.08 | 1.878604 | 5.08 | 0.850 |
| 6 | 1.878604 | 5.08 | | 5.08 | 0.850 |
| 6 | 8.281396 | | 0 | 0 | 0.850 |
| 6 | 8.281396 | | 5.08 | 5.08 | 0.850 |
| 6 | 6.958604 | | 5.08 | 0 | 0.850 |
| 6 | 6.958604 | | 0 | 5.08 | 0.850 |
| 6 | | 0 | 8.281396 | 0 | 0.850 |
| 6 | | 5.08 | 8.281396 | 5.08 | 0.850 |
| 6 | | 5.08 | 6.958604 | 0 | 0.850 |
| 6 | | 0 | 6.958604 | 5.08 | 0.850 |
| 6 | | 0 | 0 | 8.281396 | 0.850 |
| 6 | | 5.08 | 5.08 | 8.281396 | 0.850 |
| 6 | | 5.08 | 0 | 6.958604 | 0.850 |
| 6 | | 0 | 5.08 | 6.958604 | 0.850 |
| | | | | | 0.00 |

### d. Simulations of the spectrum of Prussian blue

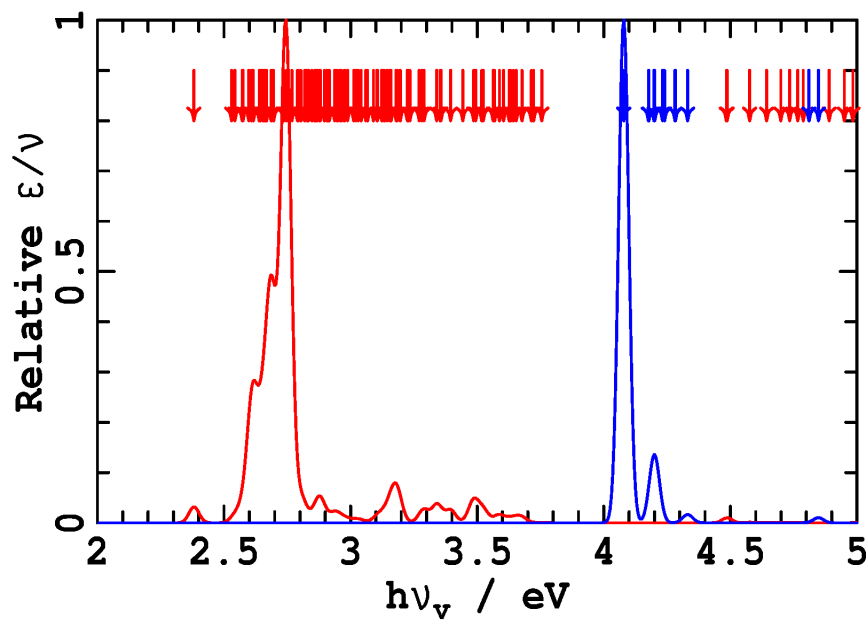

**Fig. S11.** Stick spectra depicting vertical excitation energies to excited states of Prussian blue, from CAM-B3LYP calculations of orbital energy differences and associated transition moments: red- for $Fe_3^{2+}Fe_4^{3+}(CN^-)_{18}(H_2O)_6$, blue- for $Fe^{2+}Fe^{3+}(CN^-)_6 \cdot K^+$; Gaussian resolution 10 meV. Arrows indicate calculated vertical excitation energies. These calculations neglect exciton-binding energies and exciton couplings that significantly reduce transition energies, effects that are included in Fig. 2b and Table 2 of the manuscript.

# S8. Additional references

## S9. Coordinates for model compounds and solids

```
1. SAC-CI [Fe(II)(CN)5-CN-Fe(III)(CN)5]6- staggered C4v low-spin 2E
26     0.000000     0.000000    -2.640100
26     0.000000     0.000000     2.634763
 7    -2.228070    -2.228070    -2.684893
 6     0.000000     2.125376     2.631454
 7    -2.228070     2.228070    -2.684893
 7     0.000000     0.000000    -5.792328
 6    -1.412651    -1.412651    -2.658124
 6    -1.412651     1.412651    -2.658124
 6     0.000000     0.000000    -4.640721
 6     0.000000     0.000000     4.811405
 6     2.125376     0.000000     2.631454
 7     0.000000     0.000000     0.494195
 7     0.000000     0.000000     5.969746
 7     0.000000     3.283945     2.648458
 7     3.283945     0.000000     2.648458
 6     0.000000     0.000000    -0.654391
 7     2.228070    -2.228070    -2.684893
 6     1.412651    -1.412651    -2.658124
 6    -2.125376     0.000000     2.631454
 7    -3.283945     0.000000     2.648458
 7     2.228070     2.228070    -2.684893
 6     1.412651     1.412651    -2.658124
 6     0.000000    -2.125376     2.631454
 7     0.000000    -3.283945     2.648458

#P opt=z-matrix sac-ci(cationdoublet=(nstate=(0,0,1,0))),levelone,MaxR2
 Full point group                 C4V      NOp    8
     Total energy       in au =     -1261.595787
```

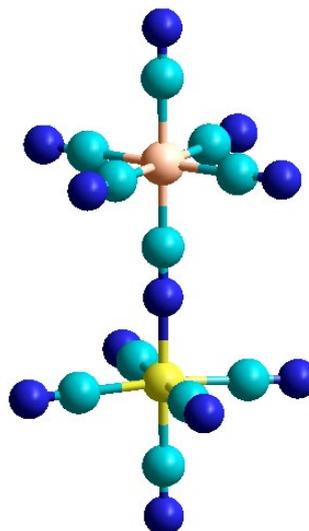

___________________________________________________________________________

```
 2.     CAM-B3LYP       [Fe(II)(CN)5-CN-
Fe(III)(NC)3(H2O)2]4- C2v high-spin 2A1
26     0.000000     0.000000    -2.460163
26     0.000000     0.000000     2.619837
 7    -3.045120     0.000000    -2.460163
 7     0.000000     0.000000    -5.526652
 7     0.000000     3.064418    -2.460163
 7     3.045120     0.000000    -2.460163
 6    -1.893090     0.000000    -2.460163
 6     0.000000     0.000000    -4.360785
 6     0.000000     1.901741    -2.460163
 6     1.893090     0.000000    -2.460163
 7     0.000000    -3.064418    -2.460163
 6     0.000000    -1.901741    -2.460163
 7     0.000000     0.000000     0.606326
 7    -2.015582     0.000000     2.619837
 7     2.015582     0.000000     2.619837
 7     0.000000     0.000000     4.633348
 6     0.000000     0.000000    -0.559541
 6    -3.178259     0.000000     2.619837
 6     3.178259     0.000000     2.619837
 6     0.000000     0.000000     5.799215
 8     0.000000     2.120512     2.619837
 1     0.000000     2.626151     3.438027
 1     0.000000     2.626151     1.801647
 8     0.000000    -2.120512     2.619837
```

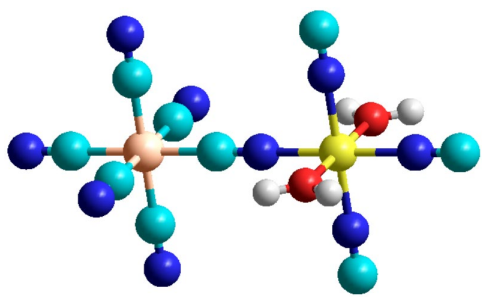

```
    1      0.000000   -2.626151    3.438027
    1      0.000000   -2.626151    1.801647

Point Charges:
 XYZ=    -5.0611     0.0000    -2.4602  Q=     0.8000
 XYZ=     0.0000     0.0000    -7.5427  Q=     0.8000
 XYZ=     0.0000     5.0804    -2.4602  Q=     0.8000
 XYZ=     5.0611     0.0000    -2.4602  Q=     0.8000
 XYZ=     0.0000    -5.0804    -2.4602  Q=     0.8000
 XYZ=     5.0713     0.0000     2.6198  Q=     0.2000
 XYZ=     0.0000     0.0000     7.6922  Q=     0.2000
 XYZ=    -5.0713     0.0000     2.6198  Q=     0.2000

 #P    cam-b3lyp/gen    td(nstate=60)    guess=read    pseudo=read    charge
pop(orbitals=50,Hi
 Full point group                    C2V         NOp    4
 SCF Done:  E(UCAM-B3LYP) =  -1235.58823195      A.U. after     18 cycles
```
______________________________________________________________________

```
 3. CAM-B3LYP [Fe(II)(CN)5-CN-Fe(III)(NC)5]6- C4v high-spin 2A1
 26    0.000000    0.000000    2.542560
 26    0.000000    0.000000   -2.537440
  7    0.000000    0.000000    5.578165
  7    3.035605    0.000000    2.542560
  7    0.000000    3.035605    2.542560
  7   -3.035605    0.000000    2.542560
  7    0.000000   -3.035605    2.542560
  7    0.000000    0.000000   -0.493045
  7    0.000000    2.044395   -2.537440
  7    2.044395    0.000000   -2.537440
  7   -2.044395    0.000000   -2.537440
  7    0.000000   -2.044395   -2.537440
  7    0.000000    0.000000   -4.581835
  6    0.000000   -1.878604    2.542560
  6   -1.878604    0.000000    2.542560
  6    0.000000    1.878604    2.542560
  6    1.878604    0.000000    2.542560
  6    0.000000    0.000000    4.421164
  6    0.000000    0.000000    0.663956
  6    0.000000   -3.201396   -2.537440
  6    0.000000    3.201396   -2.537440
  6   -3.201396    0.000000   -2.537440
  6    3.201396    0.000000   -2.537440
  6    0.000000    0.000000   -5.738836
```

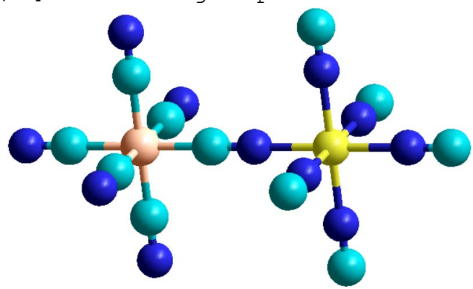

```
Point Charges:
 XYZ=     0.0000    -5.1356     2.5426  Q=     0.8000
 XYZ=     0.0000    -5.3014    -2.5374  Q=     0.2000
 XYZ=    -5.3014     0.0000    -2.5374  Q=     0.2000
 XYZ=     0.0000     0.0000     7.6782  Q=     0.8000
 XYZ=     5.3014     0.0000    -2.5374  Q=     0.2000
 XYZ=     5.1356     0.0000     2.5426  Q=     0.8000
 XYZ=    -5.1356     0.0000     2.5426  Q=     0.8000
 XYZ=     0.0000     5.1356     2.5426  Q=     0.8000
 XYZ=     0.0000     5.3014    -2.5374  Q=     0.2000
 XYZ=     0.0000     0.0000    -7.8388  Q=     0.2000

 #P cam-b3lyp/gen pseudo=read charge pop(orbitals=50,Hirshfeld)
 Full point group                    C4V         NOp    8
 SCF Done:  E(UCAM-B3LYP) =  -1268.41109283      A.U. after     30 cycles
```

4. CAM-B3LYP model tetramer C2v high-spin 2A1

```
 26    0.000000    3.592102    0.060957
 26    0.000000   -3.592102    0.060957
 26    0.000000    0.000000   -3.531145
 26    0.000000    0.000000    3.653060
  8    2.120512    0.000000    3.653060
  8   -2.120512    0.000000    3.653060
  7    0.000000   -1.424499    5.077559
  7    0.000000    1.424499    5.077559
  7    0.000000    1.424499    2.228560
  7    0.000000   -1.424499    2.228560
  7    0.000000    5.745327    2.214182
  7    0.000000   -5.745327    2.214182
  7    3.065453    3.592102    0.060957
  7    3.065453   -3.592102    0.060957
  7   -3.065453    3.592102    0.060957
  7   -3.065453   -3.592102    0.060957
  7    0.000000   -1.438877   -2.092268
  7    0.000000    1.438877   -2.092268
  7    0.000000   -5.759705   -2.106646
  7    0.000000    5.759705   -2.106646
  7    2.034880    0.000000   -3.531145
  7   -2.034880    0.000000   -3.531145
  7    0.000000    1.438877   -4.970023
  7    0.000000   -1.438877   -4.970023
  6    1.901181   -3.592102    0.060957
  6    1.901181    3.592102    0.060957
  6    0.000000   -2.247764    5.900824
  6    0.000000   -4.930719    1.399574
  6    0.000000    2.247764    5.900824
  6   -1.901181    3.592102    0.060957
  6   -1.901181   -3.592102    0.060957
  6    0.000000    2.253486   -1.277660
  6    0.000000   -2.253486   -1.277660
  6    0.000000   -4.936441   -1.283381
  6    0.000000    4.936441   -1.283381
  6    0.000000    4.930719    1.399574
  6    0.000000    2.247764    1.405296
  6    0.000000   -2.247764    1.405296
  6    3.186910    0.000000   -3.531145
  6   -3.186910    0.000000   -3.531145
  6    0.000000    2.253486   -5.784631
  6    0.000000   -2.253486   -5.784631
  1    2.626151   -0.818190    3.653060
  1    2.626151    0.818190    3.653060
  1   -2.626151    0.818190    3.653060
  1   -2.626151   -0.818190    3.653060

Point Charges:
 XYZ=   -7.1842    0.0000   -3.9523 Q=    0.8000
 XYZ=   -3.5921   -5.0800   -0.3602 Q=    0.8000
 XYZ=   -3.5921    5.0800   -0.3602 Q=    0.8000
 XYZ=   -7.1842    0.0000    3.2319 Q=    0.8000
 XYZ=    7.1842    0.0000    3.2319 Q=    0.8000
 XYZ=    3.5921    5.0800   -0.3602 Q=    0.8000
 XYZ=    7.1842    0.0000   -3.9523 Q=    0.8000
 XYZ=    3.5921   -5.0800   -0.3602 Q=    0.8000
 XYZ=    0.0000    5.0800    3.2319 Q=    0.2000
```

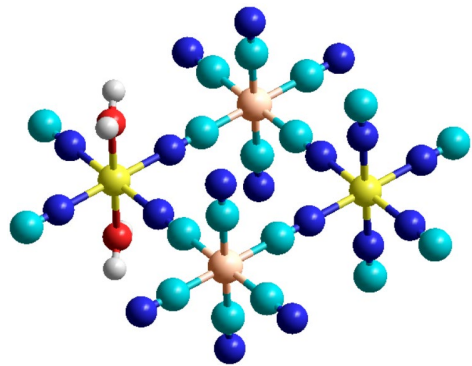

```
XYZ=     0.0000    -5.0800     3.2319  Q=       0.2000
XYZ=    -3.5921     0.0000     6.8240  Q=       0.2000
XYZ=     3.5921     0.0000     6.8240  Q=       0.2000
XYZ=    -3.5921     0.0000    -7.5444  Q=       0.2000
XYZ=     3.5921     0.0000    -7.5444  Q=       0.2000

#P cam-b3lyp/gen guess=read pseudo=read charge pop(orbitals=50,Hirshfeld)
Full point group                 C2V      NOp   4
SCF Done:  E(UCAM-B3LYP) =  -2318.15604957     A.U. after    1 cycles
```

___

5. CAM-B3LYP Prussian blue Fe(II)3Fe(III)4(CN)18(H2O)6

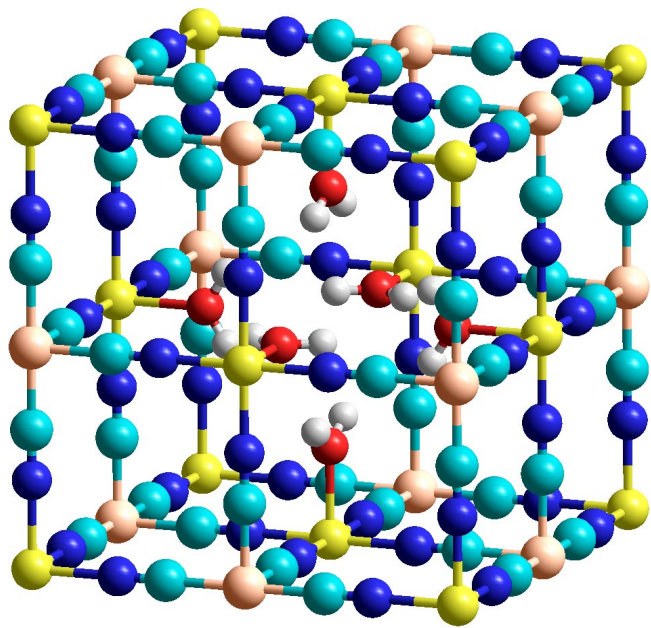

```
26    0.000000    0.000000    0.000000
26    0.000000    5.080000    5.080000
26    5.080000    5.080000    0.000000
26    5.080000    0.000000    5.080000
26    5.080000    0.000000    0.000000
26    0.000000    5.080000    0.000000
26    0.000000    0.000000    5.080000
 8    2.120512    5.080000    5.080000
 8    8.039488    5.080000    5.080000
 8    5.080000    2.120512    5.080000
 8    5.080000    8.039488    5.080000
 8    5.080000    5.080000    2.120512
 8    5.080000    5.080000    8.039488
 7    7.093511    5.080000    0.000000
 7    5.080000    7.095582    0.000000
 7    7.095582    0.000000    5.080000
 7    5.080000    0.000000    7.093511
 7    8.125120    0.000000    0.000000
 7    5.080000    3.064418    0.000000
 7    5.080000    0.000000    3.066489
 7    0.000000    7.093511    5.080000
 7    0.000000    5.080000    7.095582
 7    0.000000    8.125120    0.000000
 7    3.066489    5.080000    0.000000
 7    0.000000    5.080000    3.064418
 7    0.000000    0.000000    8.125120
 7    3.064418    0.000000    5.080000
 7    0.000000    3.066489    5.080000
 7    2.034880    0.000000    0.000000
 7    0.000000    2.034880    0.000000
 7    0.000000    0.000000    2.034880
 6    8.259378    5.080000    0.000000
 6    5.080000    8.258259    0.000000
 6    8.258259    0.000000    5.080000
 6    5.080000    0.000000    8.259378
 6    6.973090    0.000000    0.000000
 6    5.080000    1.901741    0.000000
 6    5.080000    0.000000    1.900622
 6    0.000000    8.259378    5.080000
 6    0.000000    5.080000    8.258259
 6    0.000000    6.973090    0.000000
 6    1.900622    5.080000    0.000000
 6    0.000000    5.080000    1.901741
 6    0.000000    0.000000    6.973090
 6    1.901741    0.000000    5.080000
 6    0.000000    1.900622    5.080000
 6    3.186910    0.000000    0.000000
 6    0.000000    3.186910    0.000000
```

```
   6    0.000000    0.000000    3.186910
   1    2.626151    4.261810    5.080000
   1    2.626151    5.898190    5.080000
   1    7.533849    4.261810    5.080000
   1    7.533849    5.898190    5.080000
   1    5.080000    2.626151    4.261810
   1    5.080000    2.626151    5.898190
   1    5.080000    7.533849    4.261810
   1    5.080000    7.533849    5.898190
   1    4.261810    5.080000    2.626151
   1    5.898190    5.080000    2.626151
   1    4.261810    5.080000    7.533849
   1    5.898190    5.080000    7.533849

E= -737.4363 eV
box:  10.160000 0 0 0  10.160000 0 0 0  10.160000
Kpoints:  2  2  2
basis: NGX=  50 NGY=  50 NGZ=  50 NGXF= 100 NGYF= 100 NGZF= 100
```
_____________________________________________________________________

```
   6. CAM-B3LYP original Prussian blue framework Fe(II)Fe(III)(CN)6.K
  26    5.080000    0.000000    0.000000
  26    0.000000    5.080000    0.000000
  26    0.000000    0.000000    5.080000
  26    5.080000    5.080000    5.080000
  26    0.000000    0.000000    0.000000
  26    5.080000    5.080000    0.000000
  26    5.080000    0.000000    5.080000
  26    0.000000    5.080000    5.080000
  19    2.540000    2.540000    2.540000
  19    2.540000    7.620000    7.620000
  19    7.620000    2.540000    7.620000
  19    7.620000    7.620000    2.540000
   7    3.035605    0.000000    0.000000
   7    0.000000    3.035605    0.000000
   7    5.080000    2.044395    0.000000
   7    2.044395    5.080000    0.000000
   7    0.000000    0.000000    3.035605
   7    5.080000    5.080000    3.035605
   7    5.080000    0.000000    2.044395
   7    0.000000    5.080000    2.044395
   7    2.044395    0.000000    5.080000
   7    0.000000    2.044395    5.080000
   7    5.080000    3.035605    5.080000
   7    3.035605    5.080000    5.080000
   7    7.124395    0.000000    0.000000
   7    7.124395    5.080000    5.080000
   7    8.115605    5.080000    0.000000
   7    8.115605    0.000000    5.080000
   7    0.000000    7.124395    0.000000
   7    5.080000    7.124395    5.080000
   7    5.080000    8.115605    0.000000
   7    0.000000    8.115605    5.080000
   7    0.000000    0.000000    7.124395
   7    5.080000    5.080000    7.124395
   7    5.080000    0.000000    8.115605
   7    0.000000    5.080000    8.115605
   6    1.878604    0.000000    0.000000
   6    0.000000    1.878604    0.000000
   6    5.080000    3.201396    0.000000
```

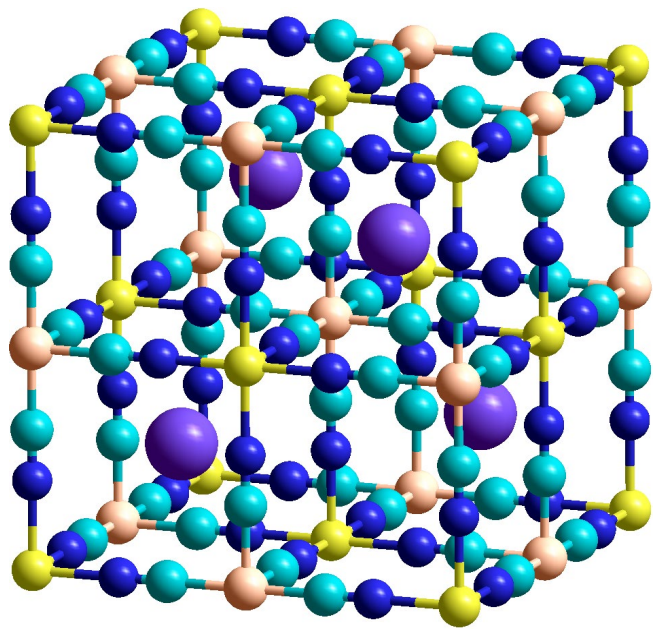

```
6      3.201396      5.080000      0.000000
6      0.000000      0.000000      1.878604
6      5.080000      5.080000      1.878604
6      5.080000      0.000000      3.201396
6      0.000000      5.080000      3.201396
6      3.201396      0.000000      5.080000
6      0.000000      3.201396      5.080000
6      5.080000      1.878604      5.080000
6      1.878604      5.080000      5.080000
6      8.281396      0.000000      0.000000
6      8.281396      5.080000      5.080000
6      6.958604      5.080000      0.000000
6      6.958604      0.000000      5.080000
6      0.000000      8.281396      0.000000
6      5.080000      8.281396      5.080000
6      5.080000      6.958604      0.000000
6      0.000000      6.958604      5.080000
6      0.000000      0.000000      8.281396
6      5.080000      5.080000      8.281396
6      5.080000      0.000000      6.958604
6      0.000000      5.080000      6.958604
```

E= -744.2397 eV  
box:   10.160000 0 0 0   10.160000 0 0 0   10.160000  
Kpoints:   2   2   2  
basis: NGX=   50  NGY=   50  NGZ=   50  NGXF= 100  NGYF= 100  NGZF= 100  

________________________________________________________________